\documentclass[a4paper,11pt]{article}
\usepackage{ifpdf} 
\ifpdf
\pdfoutput=1
\fi

\usepackage{jcappub}
\usepackage{graphicx}
\graphicspath{{./}}
\usepackage{dcolumn}
\usepackage{amssymb,amsmath,bm}
\usepackage{color}
\usepackage[dvipsnames]{xcolor}
\usepackage{xfrac}
\usepackage{aas_macros}
\usepackage{mathrsfs}
\usepackage{bbold}
\usepackage{subcaption}
\usepackage{rotating}
\usepackage{chngcntr}
\usepackage[T1]{fontenc} 
\usepackage{natbib}
\newcommand{\nv}{\hat{\boldsymbol{\theta}}}
\newcommand{\bv}{\boldsymbol{\beta}}

\newcommand{\planck}{{\sl Planck}}

\newcommand{\oxford}{Department of Physics, University of Oxford, Denys Wilkinson Building, Keble Road, Oxford OX1 3RH, United Kingdom}
\newcommand{\ipmu}{Kavli Institute for the Physics and Mathematics of the Universe (Kavli IPMU, WPI), UTIAS, The University of Tokyo, Kashiwa, Chiba 277-8583, Japan}

\definecolor{internationalkleinblue}{rgb}{0.0, 0.18, 0.65}
\hypersetup{urlcolor=internationalkleinblue, linkcolor=internationalkleinblue, citecolor=internationalkleinblue}

\title{A hybrid map-$C_\ell$ component separation method for primordial CMB $B$-mode searches}
\author[1,2]{S. Azzoni,}
\author[1]{D. Alonso,}
\author[1]{M. H. Abitbol,}
\author[3]{J. Errard,}
\author[\, 4,5,6]{N.~Krachmalnicoff}

\affiliation[1]{\oxford}
\affiliation[2]{\ipmu}
\affiliation[3]{Université Paris Cité, CNRS, Astroparticule et Cosmologie, F-75013 Paris, France}
\affiliation[4]{International School for Advanced Studies (SISSA), Via Bonomea 265, 34136, Trieste, Italy}
\affiliation[5]{Institute for Fundamental Physics of the Universe (IFPU), Via Beirut 2, 34151 Grignano (TS), Italy}
\affiliation[6]{National Institute for Nuclear Physics (INFN), Sezione di Trieste, Via Valerio 2, I-34127 Trieste, Italy}
\emailAdd{susanna.azzoni@physics.ox.ac.uk}

\abstract{The observation of the polarised emission from the Cosmic Microwave Background (CMB) from future ground-based and satellite-borne experiments holds the promise of indirectly detecting the elusive signal from primordial tensor fluctuations in the form of large-scale $B$-mode polarisation. Doing so, however, requires an accurate and robust separation of the signal from polarised Galactic foregrounds. We present a component separation method for multi-frequency CMB observations that combines some of the advantages of map-based and power-spectrum-based techniques, and which is direcly applicable to data in the presence of realistic foregrounds and instrumental noise. We demonstrate that the method is able to reduce the contamination from Galactic foregrounds below an equivalent tensor-to-scalar ratio $r_{\rm FG}\lesssim5\times10^{-4}$, as required for next-generation observatories, for a wide range of foreground models with varying degrees of complexity. This bias reduction is associated with a mild $\sim20-30\%$ increase in the final statistical uncertainties, and holds for large sky areas, and for experiments targeting both the reionisation and recombination bumps in the $B$-mode power spectrum.}

\begin{document}
\maketitle
\flushbottom

\section{Introduction}\label{sec:intro}
  The search for $B$-mode polarization of a primordial origin in the Cosmic Microwave Background (CMB) is currently one of the most compelling fundamental science cases for cosmology. Within the inflationary paradigm, $B$-mode polarisation can be generated via Thomson scattering in the early-Universe in the presence of tensor perturbations of the spacetime metric. In this context, a detection of the $B$-mode signal would therefore imply measuring the amplitude of primordial gravitational waves (through the so-called tensor-to-scalar ratio parameter $r$), which itself contains information about the details of the physical processes giving rise to inflation \citep{1997PhRvL..78.1861L}. Thus, if detected with sufficient sensitivity, this signal would allow us to rule out several families of inflationary and non-inflationary models predicting small (or no) tensor fluctuations. A sufficiently tight upper bound, at the level of $r\lesssim10^{-3}$, would also allow us to rule out important  regions of model space, such as the so-called $R^2$ models, often based on quantum corrections to gravity \cite{1979JETPL..30..682S,2008PhLB..659..703B}, which predict $r\sim1/N^2\sim10^{-3}$, where $N$ is the number of $e$-folds inflation lasts for.

  This has motivated the design of several next-generation experiments, both ground-based and space-borne, which will make precise measurements of the polarised CMB sky over a large range of frequencies in the next decade. Some of these facilities, such as the BICEP Array \cite{2016PhRvL.116c1302B}, or the Simons Observatory \cite{2019JCAP...02..056A}, have just started taking data (or will do so within the next couple of years), and current bounds on $r$ will see a sharp improvement before the end of the decade. Due to atmospheric noise and other systematics, which contaminate the largest angular scales, these ground-based experiments will be able to constrain $B$-modes on intermediate, degree-sized scales ($\ell\sim100$) through the so-called ``recombination bump'' of the CMB power spectrum. The LiteBIRD space mission \cite{hazumi2019litebird, 2020JLTP..199.1107S, 2022arXiv220202773L}, on the other hand, will start taking data in 2029, and will be able to reach tigher constraints on $r$ by targetting both the recombination and reionization bumps, the latter of which occurs at larger scales ($\ell\lesssim10$).

  Arguably the largest source of systematic uncertainty for both space and ground experiments will be contamination by the astrophysical $B$-mode signal from the Milky Way. This polarized Galactic emission is dominated at low frequencies ($\nu\lesssim60\,{\rm GHz}$) by synchrotron, while polarized thermal dust dominates at higher frequencies \cite{2014PhRvL.112x1101B,2020A&A...641A...4P}. The combination of both dominates the CMB polarization signal on the scales of interest by at least a factor $\sim5$ \cite{2018A&A...618A.166K,2020A&A...641A...4P} over the whole frequency range. Devising methods to separate this contamination from the CMB signal in a robust and efficient manner, in the presence of practical imperfections in the data taken by real instruments, is therefore of central importance in this quest.

  A successful strategy, followed by various ground-based CMB experiments, is the so-called ``multi-frequency power spectrum likelihood'' approach \cite{2013JCAP...07..025D,2015PhRvL.114j1301B,2018PhRvL.121v1301B,2021PhRvL.127o1301A, 2020arXiv200707289C}. In these \emph{$C_\ell$-based methods}, the data vector is composed of all the auto- and cross-power spectra (in this case $B$-power spectra) between sky maps at different frequencies, which is then fitted to a model that incorporates the correlations between all sky components, Galactic and cosmological. Current state-of-the-art constraints from the BICEP-{\sl Keck} collaboration have been obtained using this methodology. In its simplest incarnation, the method assumes foregrounds are perfectly correlated in frequency (i.e. they have a homogeneous spectrum across the observed sky), and a functional form is assumed for the power spectrum of the foreground amplitudes (generally in the form of a power law). The model is then characterised by a set of constant spectral parameters (e.g. spectral indices, dust temperature), and by the parameters controlling the amplitude power spectra. Spatial variation of spectral parameters, which is physically expected, should however lead to an imperfect correlation between foregrounds at different frequencies. The simplest way to account for this is by allowing for a simple scale-independent frequency decorrelation parameter, as would be the case for uncorrelated spatial variations \cite{2017A&A...603A..62V,2021PhRvL.127o1301A}. Recently a generalisation of this approach was proposed by \cite{Azzoni_2021}, using a minimal moment expansion of the foreground SEDs to account for both the amplitude and scale dependence of this effect within a consistent physical model. The method was shown to significantly improve over the vanilla approach, being able to mitigate the foreground bias for realistic sky models over areas much larger than that covered by BICEP-{\sl Keck}. This minimal expansion can be fully generalised, including all potential correlations between different spectral parameter variations and amplitudes, and avoiding any assumptions about their non-Gaussianity. This has been shown \cite{2021MNRAS.503.2478R,2022A&A...660A.111V} to yield unbiased results given a sufficiently large number of frequency channels. This methodology suffers from two main shortcomings: first, it implicitly treats foregrounds as statistically isotropic fields by compressing their information into the power spectrum, ignoring couplings between different angular modes, and approximating the power spectrum covariance as that of a Gaussian field. Secondly, by adopting a simple functional form for the power spectrum of the foreground amplitudes, which are the dominant sky components, the model may lead to biases if the real sky deviates from this functional form on the range of scales explored. This is particularly relevant for experiments targeting the low-$\ell$ regime, but it may be important at any scale given enough sensitivity.

  At the other end of the spectrum, parametric \emph{map-based methods} aim to describe the sky at the map level, separating the CMB component and using it to constrain cosmological parameters \cite{2008ApJ...676...10E, 2009ApJ...701.1804D,2016PhRvD..94h3526S,2018ApJ...853..127H, 2019arXiv191209567M, 2020A&A...641A...4P, 2022arXiv220714213M, 2022arXiv220501049V}. Note that, while these methods mostly exploit the frequency information to separate components at the map level, recently studies have been carried out on techniques to separate out Galactic foreground purely based on their spatial statistics \citep{2018MNRAS.479.5577P,2022MNRAS.510L...1J}. Map-based methods have several advantages: first, they are manifestly optimal for any specified sky model\footnote{Although $C_\ell$ methods will always achieve equivalent constraints if all components are Gaussian.}. Secondly, since all components are modelled at the map level, it is natural and simple to account for spatial variations in the foreground spectra. Finally, foreground amplitudes are reconstructed in each pixel without assuming a specific model for their spatial correlations, thus avoiding potential biases inherent in assuming a functional form for the foreground power spectra. This approach is not without shortcomings, however: first, while it is easy to incorporate spatially varying spectra in the model, controlling the scales over which spectral parameters vary is not straightforward. The most conservative approach, allowing for uncorrelated variations over all pixels in the map, can lead to significant degradation in the final cosmological constraints, especially if only a small number of frequency channels are available \citep{2016PhRvD..94h3526S,2017PhRvD..95d3504A,2019PhRvD..99d3529E}. Several methods have been proposed to select larger regions over which spectral parameters can be held constant or develop strong correlations, leading to rich and complex Bayesian forward modelling schemes \citep{2016PhRvD..94h3526S,2020MNRAS.496.4383G,2022MNRAS.511.2052P}. The computational complexity of the resulting frameworks, particularly if simultaneously sampling the CMB power spectra, is another disadvantage of these methods. While computing time is often a fair price to pay for improved or more robust constraints, it is often necessary to carry out multiple analyses of the same data before a clear understanding of all relevant sources of systematic uncertainty is achieved. Thus, fast, lightweight methods are extremely useful when confronted with new data. Finally, incorporating the impact of real-world imperfections in the data (e.g. filtering, gain or polarization angle variations) is often not straightforward in map-based methods, or may increase their computational complexity significantly.

  In this paper, we explore an alternative method combining features from map-based and $C_\ell$-based analyses. In this hybrid approach, foreground amplitudes are determined at the map level, free of assumptions regarding their statistical correlations, assuming homogeneous spectral properties. We then subtract the contribution from these foregrounds from the data, and model the residual foreground contamination (due to imperfect subtraction or spatial SED variations), at the power spectrum level. The methods is thus robust to most assumptions (Gaussianity, power spectrum form) about the dominant foreground amplitudes, while limiting the degradation in the final constraints caused by spectral index variations by imposing those assumptions on the smaller foreground residuals, while retaining information about their frequency dependence. We will study the applicability of this method for next-generation ground-based $B$-mode experiments targetting large sky areas ($f_{\rm sky}\gtrsim 0.1$) in the presence of inhomogeneous, correlated instrumental noise, for foreground models with varying degrees of complexity. We will also study the potential limitations of the method when employed on larger sky fractions, with consequently larger foreground contamination, quantifying its applicability for future space-based missions.

  The paper is structured as follows. Section \ref{sec:theory} presents the method in detail. The simulations used to validate it are described in Section \ref{sec:sims}, including the instrumental setup assumed. The method is then validated in Section \ref{sec:results}, where we explore its performance as a function of foreground complexity and sky fraction. The main conclusions of this study are summarised in Section \ref{sec:conclusion}.

\section{Methodology}\label{sec:theory}
  The component separation method proposed here consists of two stages:
  \begin{enumerate}
    \item We find a set of well-educated constant spectral indices $\beta_{\rm S}$ and $\beta_{\rm D}$ and clean out the spatially-constant part of the foregrounds at the map-level using the methods described in \cite{2009MNRAS.392..216S,2010MNRAS.408.2319S,2016PhRvD..94h3526S,2017PhRvD..95d3504A,2020MNRAS.496.4383G}. In practice, this step returns a filter ${\sf Q}$ that de-projects the linear combinations of the data that follow the best-fit SEDs of the foreground sources under consideration (dust and synchrotron in this case). This step is described in Section  \ref{ssec:theory.step1}.
    \item We use the results of the previous step to measure the auto- and cross-power spectra of the map residuals at all frequencies after removing the constant-SED foreground components. Then, under the assumption that the residual foreground contamination is small, we can model its contribution to the multi-frequency power spectrum assuming a relatively simple model, with the frequency dependence given by the derivatives of the foreground SEDs with respect to the spectral indices. As done in \cite{Azzoni_2021}, we assume a power law model for the scale dependence of the foreground residual power spectra. This step is described in Section \ref{ssec:theory.step2}.
  \end{enumerate}

  \subsection{Map-domain cleaning}\label{ssec:theory.step1}
    \subsubsection{Map-level parametrization}\label{sssec:theory.step1.map_par}
      At the map level, we model the sky signal using matrix notation as:
      \begin{equation}\label{eq:linear}
        {\bf m} = {\sf S}\,{\bf T}+{\bf n}.
      \end{equation}
      where
      \begin{itemize}
        \item[--] ${\bf m}$ is the map vector containing the measured polarized sky signal $(Q_\nu(\nv),U_\nu(\nv))$ at angular coordinates $\nv$ in each observing frequency $\nu$.
        \item[--] ${\sf S}$ is the so-called mixing matrix, with dimension $N_\nu\times N_c$, where $N_c$ is the number of components, and $N_\nu$ is the number of frequencies. This matrix encodes the emission law $S_\nu^c(\beta_c(\nv))$ of each component. We assume the same SEDs for both $Q$ and $U$.
        \item[--] ${\bf T}$ is a vector of size $N_c$ containing the amplitude of each component for a given pixel. 
        \item[--] ${\bf n}$ is a vector containing the instrumental noise at different frequencies for a given pixel. Its covariance matrix will be labelled ${\sf N}\equiv\langle {\bf n}{\bf n}^T\rangle$.
      \end{itemize}
      Note that, implicitly, we have assumed that all quantities in Eq. \ref{eq:linear}, including the foreground spectra, depend on sky coordinates $\nv$.

      We consider three components, with the following spectral dependence:
      \begin{enumerate}
        \item {\bf Thermal dust:} dust grains in the interstellar medium are heated by stellar UV radiation, producing emission on microwave frequencies. The alignment of elongated dust grains with the Galactic magnetic field produces a linear polarization perpendicular to the to both the magnetic field and the direction of propagation, making dust the most relevant foreground for $B$-mode searches on frequencies $\nu\gtrsim100\,{\rm GHz}$. Thermal dust emission is thought to be well-characterized by a modified black-body spectrum of the form
        \begin{equation}\label{eq:sed-dust}
          S_\nu^{\rm D} = \left(\frac{\nu}{\nu^{\rm D}_0}\right)^{\beta_{\rm D}}\frac{B_\nu(\Theta_{\rm D})}{B_{\nu^{\rm D}_0}(\Theta_{\rm D})},
        \end{equation}
        where $\beta_{\rm D}$ and $\Theta_{\rm D}$ are the dust spectral index and temperature.
        \item {\bf Synchrotron:} Galactic synchrotron emission is caused by the interaction of high-energy cosmic ray electrons with the Galactic magnetic field. Synchrotron is strongly polarised, and is characterised by a smooth power-law spectrum tracing the energy distribution of cosmic ray electrons. The synchrotron spectrum used here is therefore            \begin{equation}\label{eq:sed-sync}
          S_\nu^{\rm S}=\left(\frac{\nu}{\nu_0^{\rm S}}\right)^{\beta_{\rm S}},
        \end{equation}
        where $\beta_{\rm S}$ is the synchrotron spectral index.
        \item {\bf CMB:} the spectrum of CMB temperature anisotropies in antenna temperature units is
        \begin{equation}
          S_\nu^{\rm CMB}=e^x\left(\frac{x}{e^x-1}\right)^2,\hspace{12pt}x=\frac{h\nu}{k_B \Theta_{\rm CMB}},
        \end{equation}
        where $h$ is the Planck constant, $k_B$ is the Boltzmann constant, and $\Theta_{\rm CMB}=2.7255\,{\rm K}$ is the CMB monopole temperature~\cite{fixsen2009}. The CMB spectrum is isotropic and is not normalized at any pivot frequency.
      \end{enumerate}

      We will consider spatial variations in $\beta_{\rm D}$, which takes values $\beta_{\rm D}\sim1.6$. The restricted frequency range available to most ground-based experiments, including the Simons Observatory ($\nu\lesssim280\,{\rm GHz}$) makes $B$-mode studies almost insensitive to the value of $\Theta_{\rm D}$, and therefore we fix it to $\Theta_{\rm D}=19.6\,{\rm K}$ here. $\beta_{\rm S}$ is the synchrotron spectral index, which takes values $\beta_{\rm S}\sim-3$.

      We can separate the spectra ${\sf S}$ into two parts: a spatially-independent ``mean'' $\bar{\sf S}$ and a perturbation $\Delta{\sf S}$, such that Eq. \ref{eq:linear} becomes
      \begin{equation}\label{eq:map model}
        {\bf m} = \bar{\sf S}{\bf T}+\Delta{\sf S}{\bf T}+{\bf n}.
      \end{equation}
      In the first term, $\bar{\sf S}$ represents our best guess of the mean spectra of all components. Here, we will determine best-fit constant spectral indices $(\beta_{\rm S},\beta_{\rm D})$ using the map-based spectral likelihood described in Section \ref{sssec:theory.step1.like}, and $\bar{\sf S}$ is given by evaluating the spectra above at these best-fit values. The residual $\Delta{\sf S}$ includes the spatially-dependent part of the foreground spectra, as well as any difference between the estimated best-fit spectral indices and their true spatial average. The subsequent treatment of $\Delta {\sf S}$ does not change as long as these differences are small.

    \subsubsection{Map-level spectral likelihood}\label{sssec:theory.step1.like}
      The procedure to obtain the best-fit spectral indices at the map level follows the methods described in \cite{2009MNRAS.392..216S,2016PhRvD..94h3526S,2017PhRvD..95d3504A}. Given the model in Eq. \ref{eq:linear}, the posterior marginalised distribution for $\bv\equiv(\beta_{\rm D},\beta_{\rm S})$ is given by
      \begin{equation}\label{eq:map-likelihood}
        -2\ln{\mathcal{L}_\beta} = \left(\bar{\sf S}^T{\sf N}^{-1}{\bf m}\right)^T\left(\bar{\sf S}^T{\sf N}^{-1}\bar{\sf S}\right)^{-1}\left(\bar{\sf S}^T{\sf N}^{-1}{\bf m}\right),
      \end{equation}
      where ${\sf N}$ is the noise covariance matrix. As described in \cite{2017PhRvD..95d3504A}, this results from analytically marginalising over the linear amplitude parameters ${\bf T}$, and imposing a Jeffreys-like prior that corrects for volume effects in the spectral index parameters due to their non-linear nature in the model.
      
      We obtain the best-fit spectral indices $\bv_{\rm BF}$, defining $\bar{\sf S}$, as the values that maximize this likelihood, $\bv_{\rm BF}\equiv{\rm argmax}(\mathcal{L}_\beta)$, assuming that they are homogeneous across the sky. Furthermore, we use a Fisher-matrix approach to estimate the uncertainty on the spectral parameters. The inverse covariance of $\bv_{\rm BF}$ is therefore approximated as the Hessian of the log-posterior at the best fit:
      \begin{equation}
        ({\rm Cov}_{\bv})^{-1}_{ij}\simeq-\left.\frac{\partial^{2} \ln \mathcal{L}_\beta}{\partial \beta_{i} \partial \beta_{j}}\right|_{\bv=\bv_{\rm BF}}
      \end{equation}
      This approximation is accurate for sufficiently constraining data \citep{2011PhRvD..84f9907E,2016JCAP...03..052E} (as is the case of SO), and allows us to avoid sampling the full two-dimensional parameter space.

      A number of technical aspects regarding the implementation of this method in practice must be noted at this stage. First, we estimate that the data is provided in the form of splits, i.e. a set of maps containing the same sky but different noise realisations (e.g. corresponding to observations at different times). We use differences between these splits, containing only noise, to estimate the noise covariance matrix ${\sf N}$. Moreover, we assume, for simplicity, that the covariance matrix is diagonal in pixel space (i.e. noise different pixels are uncorrelated). In this paper we also assume that the noise is diagonal in frequency space, although accounting for off-diagonal correlations would pose no additional difficulty to the method. It is worth emphasizing that, even though our implementation of the method assumes uncorrelated noise for computational efficiency, this is never the case in real observations, nor in the simulations used here. As we will see, this does not spoil the validity of the method, since this first step only requires us to find spectral index values that are sufficiently close to the spatial mean over the observed footprint. Note that we do account for inhomogeneity in the noise component, using the hits count map. After calculating ${\sf N}$, we coadd all data splits into a single set of maps, which we then process as described above to find $\bar{\sf S}$.

    \subsubsection{Map-level residuals}\label{sssec:theory.step1.res}
      After determining the best-fit spectral indices and the corresponding mixing matrix $\bar{\sf S}$, the corresponding best-fit amplitudes can be found analytically by solving Eq. \ref{eq:map model} as a least-squared problem:
      \begin{equation}\label{eq:Tbf}
        {\bf T}_{\rm BF}=\left(\bar{\sf S}^T{\sf N}^{-1}\bar{\sf S}\right)^{-1}\bar{\sf S}^T{\sf N}^{-1}{\bf m}.
      \end{equation}
      It is worth emphasizing that, at this stage, we have not imposed any priors on the statistical correlations of foreground amplitudes, and instead treated them as an independent parameter in each pixel. This is one of the key differences with the usual power-spectrum-based methods, where a functional form for the foreground power spectra is often assumed.

      From ${\bf T}_{\rm BF}$ in Eq.~\ref{eq:Tbf} we can get a best-fit estimate of the foreground contribution to our observations as ${\bf m}_{\rm BF}^{\rm FG}=\bar{\sf S}{\sf P}{\bf T}_{\rm BF}$, where ${\sf P}\equiv{\rm diag}(1, 1, ..., 1, 0)$ is an $N_c\times N_c$ diagonal matrix that selects only the non-CMB components of ${\bf T}_{\rm BF}$ (we have assumed the CMB to be the last component). We can then subtract ${\bf m}^{\rm FG}_{\rm BF}$ from the data to obtain the residual maps:
      \begin{equation}\label{eq:res1}
        {\bf r}\equiv{\bf m}-{\bf m}^{\rm FG}_{\rm BF}={\sf Q}{\bf m},
      \end{equation}
      where the matrix ${\sf Q}$ is
      \begin{equation}\label{eq:Q}
        {\sf Q}\equiv\mathbb{1}-\bar{\sf S}{\sf P}\left(\bar{\sf S}^T{\sf N}^{-1}\bar{\sf S}\right)^{-1}\bar{\sf S}^T{\sf N}^{-1}.
      \end{equation}
      The matrix ${\sf Q}$ is a projector (${\sf Q}^2={\sf Q}$), and satisfies the nice property that ${\sf Q}\bar{\sf S}=\bar{\sf S}(\mathbb{1}-{\sf P})$, where $\mathbb{1}-{\sf P}$ selects only the CMB component. Note also that, since the spatial and frequency dependence of ${\sf N}$ is factorisable, ${\sf Q}$ is constant across the sky. Finally, it is straightforward to show that ${\rm Tr}({\sf Q})=N_\nu-N_c+1$. Thus, in our case, with $N_c=3$, ${\sf Q}$ projects the the $N_\nu$-dimensional dataset onto a subspace of dimension $N_\nu-2$, eliminating two independent modes corresponding to synchrotron and dust assuming constant spectral indices (i.e. ${\sf Q}\,\bar{\sf S}^{\rm D}={\sf Q}\,\bar{\sf S}^{\rm S}=0$).

       Substituting our model for ${\bf m}$ (Eq.~\ref{eq:linear}) into Eq.~\ref{eq:res1}, and using the properties of ${\sf Q}$, we obtain
      \begin{align}
        {\bf r}&={\sf Q}\left(\bar{\sf S}{\bf T}+\Delta{\sf S}{\bf T}+{\bf n}\right),\\\label{eq:res2}
        &={\bf m}_{\rm CMB}+{\sf Q}\Delta{\sf S}{\bf T}+{\sf Q}{\bf n}, 
      \end{align}
      where $({\bf m}_{\rm CMB})_\nu(\nv)\equiv S^{\rm CMB}_\nu T_{\rm CMB}(\nv)$ is the CMB component of the maps. Since the CMB row of $\Delta{\sf S}$ is zero (given that the CMB spectrum is constant), the second term in the equation above only contains foreground sources. 

      Assuming that the fluctuations in the spectral indices with respect to their constant best-fit values are small, we can use a linear expansion of $\Delta{\sf S}$ around $\bv_{\rm BF}$ (as in \cite{Azzoni_2021}):
      \begin{equation}
        \Delta S^c_\nu(\nv) = \partial_{\beta_c}\bar{S}^c_\nu\,\delta\beta_c(\nv).
      \end{equation}
      where $\partial_{\beta_c}\equiv\partial/\partial\beta_c$, and we do not implicitly sum over $c$. Then, we can rewrite Eq.~\ref{eq:res2} as:
      \begin{equation}\label{eq:res3}
        r_\nu(\nv)=S^{\rm CMB}_\nu T_{\rm CMB}(\nv)+\tilde{S}^c_\nu\tilde{T}_c(\nv)+\tilde{n}_\nu(\nv),
      \end{equation}
      where the modified spectra and amplitudes are:
      \begin{equation}\label{eq:mod-spectra}
        \tilde{S}^c_\nu\equiv Q^{\nu'}_\nu\,\partial_{\beta_c}\bar{S}^c_{\nu'},
        \hspace{12pt}
        \tilde{T}_c(\nv)\equiv\delta\beta_c(\nv)T_c(\nv),
      \end{equation}
      and we implicitly sum over $\nu'$.

      In other words, if the spectral indices are truly constant across the sky, and if the inferred value of $\bv$ is equal to the true one, then ${\bf r}$ in Eq.~\ref{eq:res1} is a set of multi-frequency maps, each containing the same copy of the CMB. Otherwise each map in Eq.~\ref{eq:res1} contains a spectral residual, with a frequency dependence given by the first-order expansion of each component's SED. We will then model and marginalise over these residuals at the power spectrum level, as described in the next section.

  \subsection{$C_{\ell}$ likelihood on deprojected spectral modes}\label{ssec:theory.step2}
    To separate the CMB contribution at the level of the power spectrum, we use a multi-frequency power spectrum likelihood, similar to that used in the latest analysis carried out by the BICEP-{\sl Keck} collaboration \cite{2016PhRvL.116c1302B,2018PhRvL.121v1301B}. In this case, the data vector is the full matrix of cross-power spectra $C_\ell^{\nu\nu'}$ between all pairs from a set of frequency maps\footnote{The code is available at \url{https://github.com/simonsobs/BBPower}.}. The model used to describe this data vector, and the procedure used to estimate it from the data are described here.

    \subsubsection{Power spectrum model}\label{sssec:theory.step2.pred}
      Here, the map model described above gets propagated through to power spectrum. We start from a linear model of the form:
      \begin{equation}\label{eq:model_gen}
        d_\nu(\nv)={\cal S}^c_\nu\,{\cal T}_c(\nv)+n_\nu,
      \end{equation}
      where $d_\nu(\nv)$ is a set of frequency maps, ${\cal T}_c(\nv)$ is a set of component amplitude maps, and ${\cal S}^c_\nu$ is a {\sl constant} matrix characterising the spectra of the different components. Here we will explore two possible approaches:
      \begin{enumerate}
        \item In the {\bf baseline} case, $d_\nu$ is the raw set of frequency maps, and ${\cal S}^c_\nu$ and ${\cal T}_c$ are the SEDs and amplitude maps of dust, synchrotron, and the CMB.
        \item In the {\bf hybrid} case, $d_\nu$ is the set of residual maps described in the previous section, and $({\cal S}^c_\nu,{\cal T}_c)$ are given by the modified spectra and amplitudes $(\tilde{S}^c_\nu,\tilde{T}_c)$ in Eq. \ref{eq:mod-spectra}, complemented with the CMB SED and amplitude map. 
      \end{enumerate}
    
      Since we assume the mixing matrix ${\cal S}^c_\nu$ to be spatially constant, the theoretical prediction for the multi-frequency power spectrum of the map-level model in Eq. \ref{eq:model_gen} is simply given by
      \begin{equation}\label{eq:cl_lin}
        C^{\nu\nu'}_\ell=\sum_{c,c'}{\cal S}^c_\nu {\cal S}^{\nu'}_{c'}\,C^{cc'}_\ell,
      \end{equation}
      where $C^{cc'}_\ell$ is the cross-power spectrum between the amplitude maps of components $c$ and $c'$.

      Following the model used by \cite{2018PhRvL.121v1301B}, we parametrize $C_\ell^{cc'}$ for the different components described above as follows.
      \begin{itemize}
        \item Foreground auto-correlations are modelled as power-laws of the form:
        \begin{equation}\label{eq:cl_amp}
          \frac{\ell(\ell+1)}{2\pi} {\sf C}_\ell^{cc} = A_c\left(\frac{\ell}{\ell_0}\right)^{\alpha_c} ,
        \end{equation}
        with $c \in \{{\rm D},{\rm S}\}$, $\ell_0\equiv80$.
        \item The dust-synchrotron cross-correlation is parametrized through a scale-independent correlation coefficient $\epsilon_{\rm DS}$: 
        \begin{equation}
          C_\ell^{\rm DS}=~\epsilon_{\rm DS}\sqrt{C_\ell^{\rm DD}C_\ell^{\rm SS}}
        \end{equation}
        \item The CMB $B$-mode power spectrum is parametrized as
        \begin{equation}\label{eq:cl_cmb_bb}
         C_{\ell}^{\rm CMB} = A_{\rm lens}\,C^{\rm lens}_\ell+r\,\left.C^{\rm tens}_\ell\right|_{\rm{r = 1}},
        \end{equation}
        where $C^{\rm lens}_\ell$ and $C^{\rm tens}_\ell|_{\rm{r=1}}$ are templates for the $B$-mode power spectrum caused by gravitational lensing and by primordial tensor fluctuations with tensor-to-scalar ratio $r=1$ respectively. We assume no correlation between the CMB and foreground components.
      \end{itemize}

      In both the ``baseline'' and ``hybrid'' approaches, the model has a total of 9 free parameters: $\boldsymbol{\theta}=\{A_{\rm lens},r,A_{\rm D},\alpha_{\rm D},\beta_{\rm D},A_{\rm S},\alpha_{\rm S},\beta_{\rm S},\epsilon_{\rm DS}\}$. However, since the bulk of the foreground contamination is removed at the map level in the hybrid approach, we will fix $\epsilon_{\rm DS}=0$ in that case, and explore the impact of freeing up this parameter. Note that, although the best-fit $\beta_{\rm D}$ and $\beta_{\rm S}$ have been determined in the first step of the ``hybrid'' method, we allow for them to vary in the power spectrum likelihood, with a prior given by the uncertainties calculated in that step.
   
    \subsubsection{Power spectrum likelihood}\label{sssec:theory.step2.cllike}
      Given a set of measured power spectra $\hat{C}_\ell^{\nu\nu'}$, and their theoretical prediction outlined in the previous section, construct a Gaussian likelihood of the form
      \begin{equation}\label{eq:gauslike}
        \chi^2\equiv-2\ln p(\hat{\sf C}_{\ell}|\boldsymbol{\theta})=\sum_{a,b} [C^{ij}_\ell-\hat{C}^{ij}_{\ell}]\,({\sf \Sigma}^{-1})_{ij\ell,mn\ell'} [C^{mn}_{\ell'}(\theta)-\hat{C}^{mn}_{\ell'}],
      \end{equation}
      where ${\sf \Sigma}$ is the covariance matrix of the power spectrum measurements. Note that, strictly speaking, power spectra of Gaussian fields follow a Wishart distribution, which can be approximated as proposed by \cite{2008PhRvD..77j3013H}, but we verified that the Gaussian approximation is sufficiently accurate on the scales used here ($30 \leq \ell \leq 300 $). 
      This approximation becomes inaccurate on larger scales, where the small number of available modes invalidates the central limit theorem and, as quadratic functions of Gaussian fields, power spectra follow a markedly non-Gaussian distribution. To account for this effect, we employ the method of \cite{2008PhRvD..77j3013H} (``HL'' hereon) when including scales $\ell < 30$ (see Section \ref{ssec:res.large_fsky}). 
      The bias, however, may be caused by the use of the Gaussian likelihood presented in Section \ref{sssec:theory.step2.cllike}, which becomes inaccurate on scales $\ell\lesssim 30$. In fact, on larger scales 

      As described in \cite{Azzoni_2021}, we estimate the cross-correlation of the data splits introduced in Section \ref{sssec:theory.step1.map_par} in order to avoid modelling and subtracting the noise bias. We also make use of 100 Gaussian simulations, described in Section \ref{ssec:sims.gaussian}, to estimate the covariance matrix ${\sf \Sigma}$. In order to avoid numerical noise in the covariance, we set all off-diagonal elements coupling two $\ell$s separated by more than 2 bandpowers to zero. When applying our method to simulations with larger sky areas, in Section \ref{ssec:res.large_fsky}, we will approximate the covariance using the ``Knox formula'' \cite{1997ApJ...480...72K}:
      \begin{align}\label{eq:knox-covariance}
        {\sf \Sigma}_\ell^{a,b} \equiv \langle \Delta\hat{{\sf C}}_\ell^{a,b} \Delta\hat{{\sf C}}_\ell^{c,d} \rangle = \delta_{\ell \ell'} \frac{{\sf C}_\ell^{a,c}{\sf C}_\ell^{b,d} + {\sf C}_\ell^{a,d}{\sf C}_\ell^{b,c}}{(2\ell +1)\Delta\ell \, f_{sky}},
      \end{align}
      where $\Delta\ell$ is the number of multipoles in the bandpower labelled with multipole $\ell$, and $f_{\rm sky}$ is the usable sky fraction.

      We measure power spectra using bandpowers of width $\Delta\ell = 10$. By default, we will only include scales in the range $30\leq \ell\leq 300$, appropriate for a ground-based experiment, affected by atmospheric noise, targeting a $B$-mode detection from the recombination bump.

    \subsubsection{Residual $C_\ell$s and covariance}\label{sssec:theory.step2.rescl}
      In both the ``baseline'' and ``hybrid'' approaches, the parameter inference starts from the full set of cross-correlations between all raw frequency maps, $\hat{\sf C}_\ell$. In the ``baseline'' approach, these are compared directly with the model in Section \ref{sssec:theory.step2.pred}. In the ``hybrid'' approach, however, we must first construct the power spectra of the residual maps.

      The residual maps ${\bf r}$ are related to the raw maps ${\bf m}$ via Eq.~\ref{eq:res2}, where ${\sf Q}$ is the projector matrix defined in Eq.~\ref{eq:Q}, and estimated in the first step of the method from the map-level likelihood. The power spectrum of the residual maps, ${\sf C}_\ell^r$ is thus related to the power spectrum of the raw maps simply via
      \begin{align}
        {\sf C}^r_\ell = {\sf Q}\cdot {\sf C}_\ell\cdot{\sf Q}^T.
      \end{align}
      Although it would be tempting to simply apply the modelling and likelihood described in the preceding sections to $\hat{\sf C}^r_\ell$ calculated as above from the raw power spectrum measurements, this is, unfortunately, not possible. Since ${\sf Q}$ projects out the two independent modes corresponding to synchrotron and dust with constant spectral indices, the rank of ${\sf C}^r_\ell$ is $N_\nu-2$, and not $N_\nu$, and therefore has null eigenvalues. This will also be the case for the power spectrum covariance, which is therefore not invertible. This makes the Gaussian likelihood in Eq. \ref{eq:gauslike} numerically unstable.

      To solve this problem, rather than making use of pseudo-inverses, we explicitly construct a new basis for our data in the subspace defined by the image of ${\sf Q}$, and express the data and covariance in this lower-dimensional space. Since ${\sf Q}$ is a projector, and ${\rm Tr}({\sf Q})=N_\nu-2$, we can diagonalise it as
      \begin{equation}
        {\sf Q}={\sf B}\,{\sf D}\,{\sf B}^{-1},
      \end{equation}
      where ${\sf B}$ is a matrix containing the eigenvectors of ${\sf Q}$ as columns, and ${\sf D}={\rm diag}(1,...,1,0,0)$ is the eigenvalue matrix. We have arbitrarily selected the eigenvector order such that the two null eigenvectors are at the end. Let $\tilde{\sf D}$ be the matrix that results from removing the last two null rows of ${\sf D}$, and define the matrix ${\sf R}\equiv\tilde{\sf D}{\sf B}^{-1}$. ${\sf R}$ has dimensions $(N_\nu-2)\times N_\nu$, and preserves the important property of ${\sf Q}$ that ${\sf R}\bar{\sf S}^{\rm D}={\sf R}\bar{\sf S}^{\rm D}=0$. Thus ${\sf R}$ projects general $N_\nu$-dimensional vectors onto the $(N_\nu-2)$-dimensional image of ${\sf Q}$, explicitly reducing the dimensionality of the resulting vector. In order to obtain non-singular residual power spectra and covariances, we therefore use ${\sf R}$ instead of ${\sf Q}$. 

      In summary, given the $N_\nu\times N_\nu$ matrix of multi-frequency power spectra $C_\ell^{\nu\nu'}$, and its covariance matrix, we construct the $(N_\nu-2)\times(N_\nu-2)$ residual power spectrum matrix
      \begin{equation}\label{eq:r_sandwich}
        \tilde{C}^{\alpha\beta}_\ell\equiv R_{\alpha\nu}R_{\beta\nu'}C^{\nu\nu'}_\ell,
      \end{equation}
      and its covariance matrix
      \begin{align}\label{eq:r_sandwich_cov}
        {\rm Cov}\left(\widetilde{C}_\ell^{\alpha \beta},\widetilde{C}_{\ell'}^{\gamma \delta}\right) = R_{\alpha \nu_1} R_{\beta \nu_2}R_{\gamma \nu_3}R_{\delta \nu_4} \ {\rm Cov}\left({C}_\ell^{\nu_1\nu_2}{C}_\ell^{\nu_3 \nu_4}\right).
      \end{align}
      Note that the transformation in Eq. \ref{eq:r_sandwich} is applied to both the input power spectrum data, and to the theoretical prediction described in Section \ref{sssec:theory.step2.pred}.

  \subsection{Summary of the method}
    The hybrid method proposed here can be summarised as follows.
    \begin{enumerate}
      \item Starting from a set of $N_\nu$ frequency maps ${\bf m}$, we obtain the best-fit spectral index parameters $\bv_{\rm BF}$, the associated spectra, and the projector ${\sf Q}$ that removes the best-fit dust and synchrotron components from the data, using the map-level spectral likelihood of Section \ref{sssec:theory.step1.like}. We also construct the $(N_\nu-2)\times N_\nu$ matrix
      \begin{equation}
        {\sf R}\equiv\tilde{\sf D}\,{\sf B}^{-1},
      \end{equation}
      where ${\sf B}$ is the matrix of eigenvectors of ${\sf Q}$, and $\tilde{\sf D}$ selects only the rows of ${\sf B}^{-1}$ corresponding to the non-zero eigenvalues of ${\sf Q}$.
      \item We estimate the full set of power spectra between all pairs of frequencies, $\hat{C}^{\nu\nu'}_\ell$, from the same maps, and their covariance matrix.
      \item We use ${\sf R}$ to deproject the constant dust and synchrotron modes from $\hat{C}^{\nu\nu'}_\ell$ and its covariance, according to Eqs. \ref{eq:r_sandwich} and \ref{eq:r_sandwich_cov}.
      \item We calculate the theoretical prediction for these power spectra according to Eq. \ref{eq:cl_lin}, using simple power-law models to describe the scale dependence of the foreground residuals, and the derivative of the foreground SEDs to describe their frequency dependence (see Section \ref{sssec:theory.step1.res}). We apply the ${\sf R}$ matrix to this prediction using again Eq. \ref{eq:r_sandwich}.
      \item We use a Gaussian likelihood to obtain parameter constraints by comparing the data against this theoretical prediction. When inferring parameter constraints, by default we vary the CMB parameters $\{A_{\rm lens},r\}$, the foreground power spectrum parameters $\{A_c,\alpha_c\}$ ($c\in\{{\rm D},{\rm S}\}$), and the spectral indices $\{\beta_{\rm D},\beta_{\rm S}\}$, imposing a prior on the latter based on the map-level constraints.
    \end{enumerate}

    The main differences with respect to the ``baseline'' power-spectrum-based approach are that the latter skips steps 1 and 3 above, applies the power-law modelling for the full foreground power spectra (instead of the residuals), assuming constant spectral indices, and therefore does not make use of the ${\sf R}$ matrix to project onto the space of residuals. The assumption of constant spectral indices can be relaxed at the power spectrum level by implementing a moment expansion method, as done in \citep{Azzoni_2021}, or by allowing for frequency decorrelation \citep{2017A&A...603A..62V}. The main shortcomings of these purely $C_\ell$-based methods, which the hybrid approach is able to address, are the need to assume a model for the scale dependence of the foreground amplitudes, and to either ignore spectral index variations, or to assume a model for their statistics (e.g. Gaussian, uncorrelated with amplitude variations, or with a flat scale dependence). In turn, the hybrid approach only assumes a model for the scale dependence for the much smaller foreground residuals, limiting the number of parameters needed to model them. 

    It must be noted that, as described above, the method proposed is not rigorous from a statistical point of view, as it does not propagate the uncertainties of all parameters self-consistently. Instead of splitting the parameter inference into two stages (map-based and $C_\ell$-based), a more principled approach, similar to \cite{2008ApJ...676...10E}, would simultaneously sample the linear foreground and CMB amplitudes, the constant spectral indices, the power spectrum of the foreground residuals, and the CMB power spectrum parameters, in a joint map-level likelihood, effectively treating the foreground residuals as a two-dimensional Gaussian process. While fully self-consistent, this kind of approach is significantly more computationally demanding, although not intractable. Instead, we first use the maps to obtain tight constraints on the average spectral indices, and then reuse the data at the power spectrum level to marginalise over the foreground residuals. As we will show, however, in practice we find that the method proposed here is able to produce reliable constraints when applied to simulated data. The main reason for this is the fact that the information contained in the maps, for the instrumental sensitivities explored here, allows for a very precise measurement of the mean spectral indices in the first step. This first step is therefore akin to obtaining an external, high-precision measurement of $\bv$, which can be used to deproject the constant foreground amplitudes from the data, rendering the procedure fully self-consistent.

\section{Simulations}\label{sec:sims}
  In order to test the validity of the hybrid component separation method described in the previous section, we test it on a suite of sky simulations. These simulations include the most relevant sky components, as discussed in Section~\ref{sec:theory}. In all simulations, the CMB contribution was generated as a Gaussian random field drawn from the power spectrum in Eq.~\ref{eq:cl_cmb_bb}. We use fiducial values $(A_{\rm lens}=1,\,r=0)$, unless otherwise stated. 
  The simulations also incorporate the contribution from instrumental noise and the effects of a limited sky coverage as described in Section~\ref{ssec:sims.instrum}. We generate simulations of the polarized Stokes $Q$ and $U$ sky maps in six frequency bands as described in Table \ref{tab:SO}. These maps are generated using the {\tt HEALPix} pixelization scheme \cite{2005ApJ...622..759G} with resolution parameter $N_{\rm side}=256$.

  \subsection{Instrumental effects}\label{ssec:sims.instrum}
    \begin{table}[tbp]
      \centering
      \begin{tabular}{|c|c|c|c|c|}
        \hline
        Frequency & FWHM & Noise (baseline) & $\ell_{\rm knee}$ & $\alpha_{\rm knee}$\\
        (GHz) & (arcmin) & ($\mu K$-arcmin) & -- & --\\
        \hline\hline
        27 & 91 & 35 & 15 & -2.4\\
        39 & 63 & 21 & 15 & -2.4\\
        93 & 30 & 2.6 & 25 & -2.5\\
        145 & 17 & 3.3 & 25 & -3.0\\
        225 & 11 & 6.3 & 35 & -3.0\\
        280 & 9 & 16 & 40 & -3.0\\
        \hline
      \end{tabular}
      \caption{Summary of the beam Full Width at Half Maximum (FWHM) apertures, baseline and goal sensitivity levels for each band of the SO Small Aperture Telescope (SAT) from \cite{2019JCAP...02..056A}. The correlated noise power spectrum is parametrized as $N_\ell=N_{\rm white}[(\ell/\ell_{\rm knee})^{\alpha_{\rm knee}}+1]$.}\label{tab:SO}
    \end{table}
    \begin{figure}
    \centering
        \includegraphics[width=0.6\textwidth]{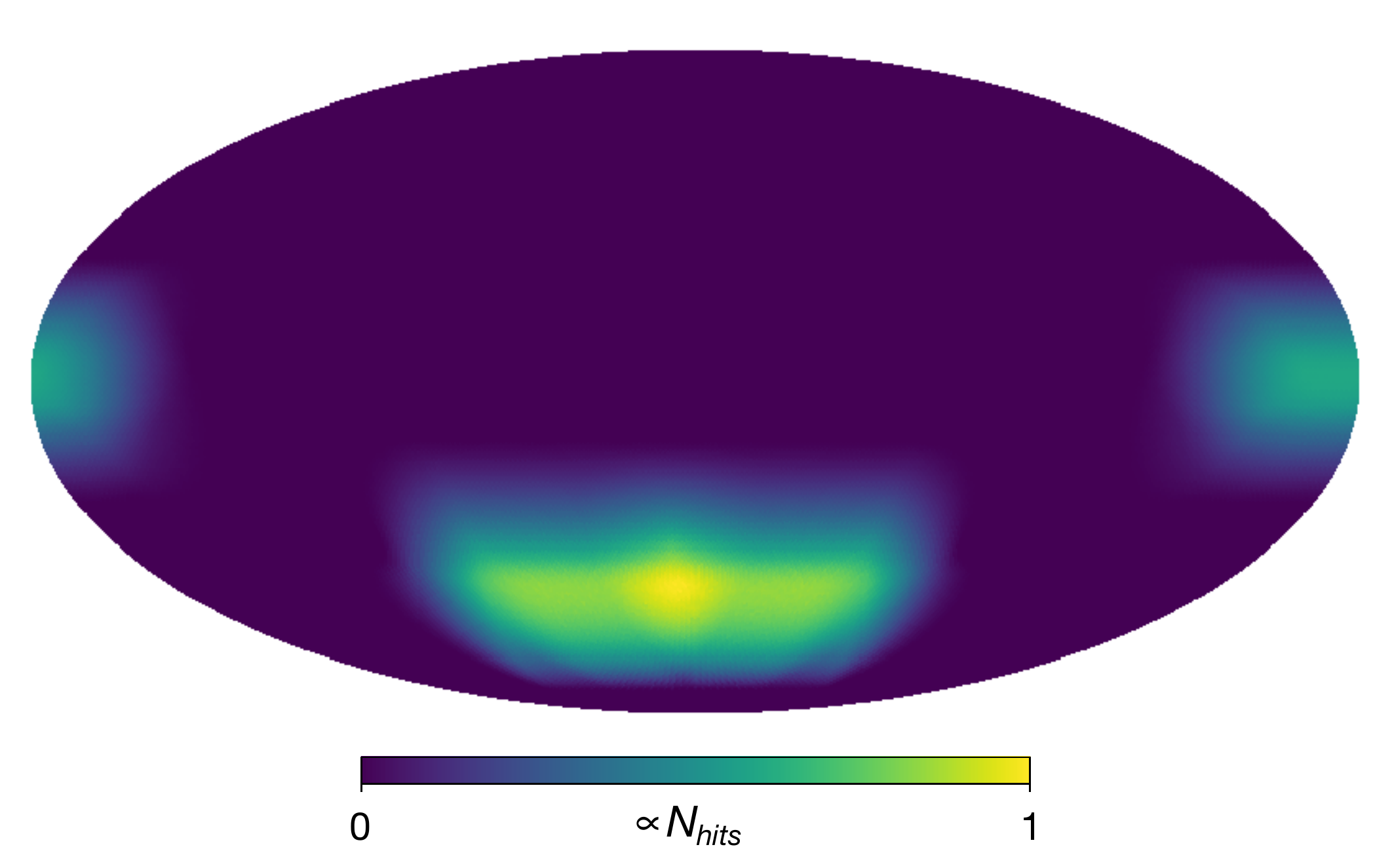}
        \caption{Sky mask used in the analysis, proportional to the map of hit counts in Equatorial coordinates used for the SO SATs.
        }\label{fig:nhits_mask}
    \end{figure}
    All the simulations produced in this work include a model of the instrumental noise which replicates the characteristics of the Simons Observatory (SO), as described in \cite{2019JCAP...02..056A} (SO19 hereon). The most relevant effects are the scale dependence and inhomogeneity of the noise, and the limitedsky coverage. 
    \begin{itemize}
      \item \textbf{Scale-dependent noise}. We model the noise using the noise calculator released with the data supplement of SO19. This uses a noise power spectrum of the form:
      \begin{equation}\label{eq:nl}
        N_\ell=N_{\rm white}\left[\left(\frac{\ell}{\ell_{\rm knee}}\right)^\alpha+1\right],
      \end{equation}
      where $N_{\rm white}$ is the white noise variance (in units of $\mu K_{\rm CMB}^2\,{\rm srad}$), while $\ell_{\rm knee}$ and $\alpha_{\rm knee}$ parametrise the scale at which $1/f$ noise starts dominating, and the steepness with which it rises on large scales, respectively. The values for these three parameters used here are listed in Table \ref{tab:SO}.
      \item \textbf{Sky footprint and inhomogeneous noise}. Except in Section~\ref{ssec:res.large_fsky}, we will assume a sky footprint that follows the hits count map provided with the SO19 data supplement, and shown in Fig.~\ref{fig:nhits_mask}. Inhomogeneous noise realisations are created by generating a homogeneous Gaussian random field with the power spectrum of Eq. \ref{eq:nl}, and then scaling the resulting noise maps with the inverse square-root of this hits map. We also use the hits map to construct a differentiable sky mask by first smoothing the hits map with a $1^\circ$ Gaussian beam, and then applying a 5$^\circ$ ``C1' apodization (see \cite{2019JCAP...02..056A}). We use this mask to compute all power spectra (except in Section \ref{ssec:res.large_fsky}) using a pseudo-$C_\ell$ algorithm with $B$-mode purification as implemented in {\tt NaMaster}\footnote{\url{https://github.com/LSSTDESC/NaMaster}} \cite{2019MNRAS.484.4127A}.
    \end{itemize}
    Note that we have not included the effects of an instrumental beam in these simulations in order to simplify the analysis. The impact of beams is straightforward to incorporate in the power spectrum analysis, and thus does not impact this method. At the map-level stage, maps should be pre-smoothed to a common beam before obtaining the best-fit spectral parameters. Again, this should not invalidate any of the results presented here. 

    As mentioned in Section \ref{ssec:theory.step1}, for each sky simulation, we generate 4 independent noise maps, each with a noise amplitude twice as large as the target sensitivity. This allows us to use these 4 realizations as independent data splits. Therefore, each simulation consists of 24 pairs of $(Q,U)$ maps, corresponding to the 6 frequency channels and 4 data splits. In the map-domain cleaning described in Section~\ref{ssec:theory.step1}, the data vector which enters the spectral likelihood in Eq.~\ref{eq:map-likelihood} is given by
    \begin{equation}
        \hat{{\bf m}} = \frac{1}{N_{\rm splits}}\sum_i^{N_{\rm splits}}({\bf s} + {\bf n}_i),
    \end{equation}
    i.e. the map of the coadded splits. When computing the multi-frequency power spectrum matrix that forms the basis of the $C_\ell$-cleaning stage, we start by computing all $BB$ auto- and cross-correlations between different pairs of maps, for a total of $300$ different $BB$ power spectra between all $N_{\rm splits}=4$ splits and $N_{\nu}=6$ frequency bands. In order to avoid modelling the noise bias as part of the likelihood described in Sec.~\ref{ssec:theory.step2}), we then generate a coadded power spectrum matrix by averaging over all power spectra combining different data splits for each frequency pair. This results in a collection of $21$ distinct coadded power spectra, which form the independent elements of the $6\times6$ multi-frequency power spectrum matrix.

  \subsection{Gaussian foreground simulations}\label{ssec:sims.gaussian}
    We generate a large suite of ``Gaussian'' foreground realizations. For these, we simulate sky maps for the amplitude (${\bf T}_c(\nv)$) of each component as Gaussian random fields governed by power spectra following the power-law model in Eq. \ref{eq:cl_amp}.

    The aim of these Gaussian simulations is twofold. First, we employ them to compute the power-spectrum covariance as described in Section \ref{sssec:theory.step2.cllike}. To estimate covariance matrices, we generate a suite of 100 Gaussian simulations with constant spectral indices.

    Secondly, we use these Gaussian simulations to study the performance of the method proposed here with different levels of spectral index variation. We parametrize this variation in terms of the standard deviation of $\delta\beta_c(\nv)$ at the pixel level. This is given in terms of the power spectrum of $\delta\beta_c$ as:
    \begin{equation}
      \sigma^2(\beta_c)=\sum_{\ell=\ell_{\rm min}}^{\ell_{\rm max}}\frac{2\ell+1}{4\pi}C_\ell^{\beta_c}.
    \end{equation}
    As in \cite{Azzoni_2021}, we model the power-spectrum of the spectral index variations $C_\ell^{\beta_c}$ as a power-law of the form
    \begin{align}\label{eq:cl_beta}
      C_\ell^{\beta_c}=B_c\left(\frac{\ell}{\ell_0}\right)^{\gamma_c},
    \end{align}
    with $\ell_0=80$ in all cases. This model is in reasonable agreement with constraints from current measurements of the synchrotron and dust spectral index \citep{2016A&A...594A...9P}. The spectral index maps are generated as Gaussian realizations of $C_\ell^{\beta_c}$ in Eq.~\ref{eq:cl_beta} with varying values for the amplitude $B_c$. Instead of varying $B_c$ directly, we generate maps of $\delta\beta_c(\nv)$ with an arbitrary amplitude and then renormalize them to enforce a given per-pixel standard deviation $\sigma({\beta_c})$. We generate simulations for $\sigma(\beta_c)=\{0,0.1,0.2,0.3\}$ for both synchrotron and dust, which covers. We fix the spectral tilt $\gamma_c$ to $(\gamma_{\rm D},\gamma_{\rm S})=(-3.5, -2.5)$. We then add the mean spectral indices $\bar{\beta}_c$ to $\delta\beta_c(\nv)$ and use the Python Sky Model software ({\tt PySM}\footnote{\href{https://github.com/bthorne93/PySM_public}{https://github.com/bthorne93/PySM\_public}}, \cite{thorne2017python}) to generate observed sky maps in the six frequency channels of Table \ref{tab:SO}.

    All the Gaussian simulations presented here use the same values for the constant spectral indices ($\bar{\beta}_{\rm D}=1.6$, $\bar{\beta}_{\rm S}=-3$) and the same pivot frequencies ($\nu_0^{\rm D}=353\,{\rm GHz}$, $\nu_0^{\rm S}=23\,{\rm GHz}$). The dust-synchrotron correlation coefficient was set to $\epsilon_{\rm SD}=0$. We use the foreground parameters that best fit the dust and synchrotron template maps used in the realistic set of simulations, described in Section~\ref{ssec:sims.pysm}, within the footprint used here:
    \begin{align}\label{eq:gauss_params}
      &A_{\rm D}=28\,\mu{\rm K}^2,\hspace{12pt} \alpha_{\rm D}=-0.16,\hspace{12pt} A_{\rm S}=1.6\,\mu{\rm K}^2,\hspace{12pt} \alpha_{\rm S}=-0.93.
    \end{align}

  \subsection{Realistic foreground simulations}\label{ssec:sims.pysm}
    \subsubsection{``\texttt{d1sm}'' simulations}  \label{sssec:sims.pysm.d1sm}
      We produce an additional set of 20 simulations with a higher level of complexity, including the foreground templates provided by {\tt PySM}. These models include ``realistic'' non-Gaussian amplitude and spectral index maps that are based on existing observations.

      Specifically, for the dust amplitude and spectral index map we adopt the templates included in the {\tt d1} model. The spatially-varying dust spectral index map in {\tt d1} is consistent with the estimate from \planck{} data using the {\tt Commander} component separation code \citep{2016A&A...594A..10P}. Similarly, we use the synchrotron amplitude map assumed by the {\tt s1} model. However, we use a modified version of the {\tt s1} model for the synchrotron spectral index $\beta_{\rm S}$. The original {\tt s1} spectral-index map was determined by combining the Haslam 408 GHz map \cite{haslam1981408, haslam1982408} with the WMAP 23 GHz map \cite{dunkley2009prospects, 2013ApJS..208...20B}. It was shown by \cite{2018A&A...618A.166K}, that the value of $\beta_{\rm S}$ and its level of variation is too low compared to the data acquired by the SPASS experiment. We therefore generate a synchrotron spectral index map by re-scaling the {\tt s1} map and extending it to smaller scales using a power-law spectral index power spectrum $C_\ell^{\beta_{\rm S}}$ consistent with the measurements of \cite{2018A&A...618A.166K}. The level of realism included in these simulations is compatible with the one used to assess the performance of forthcoming $B$-mode experiments in e.g. \cite{2019JCAP...02..056A,2019arXiv190704473A,2020arXiv200812619T}. We label the resulting set of sims ``{\tt d1sm}''. Assuming free amplitudes at the first step of our pipeline, as described in Sec.~\ref{ssec:theory.step1}, should allow us to account for the higher complexity of these models.
    
      \begin{figure}[h!t]
        \centering
        \includegraphics[width=0.9\textwidth]{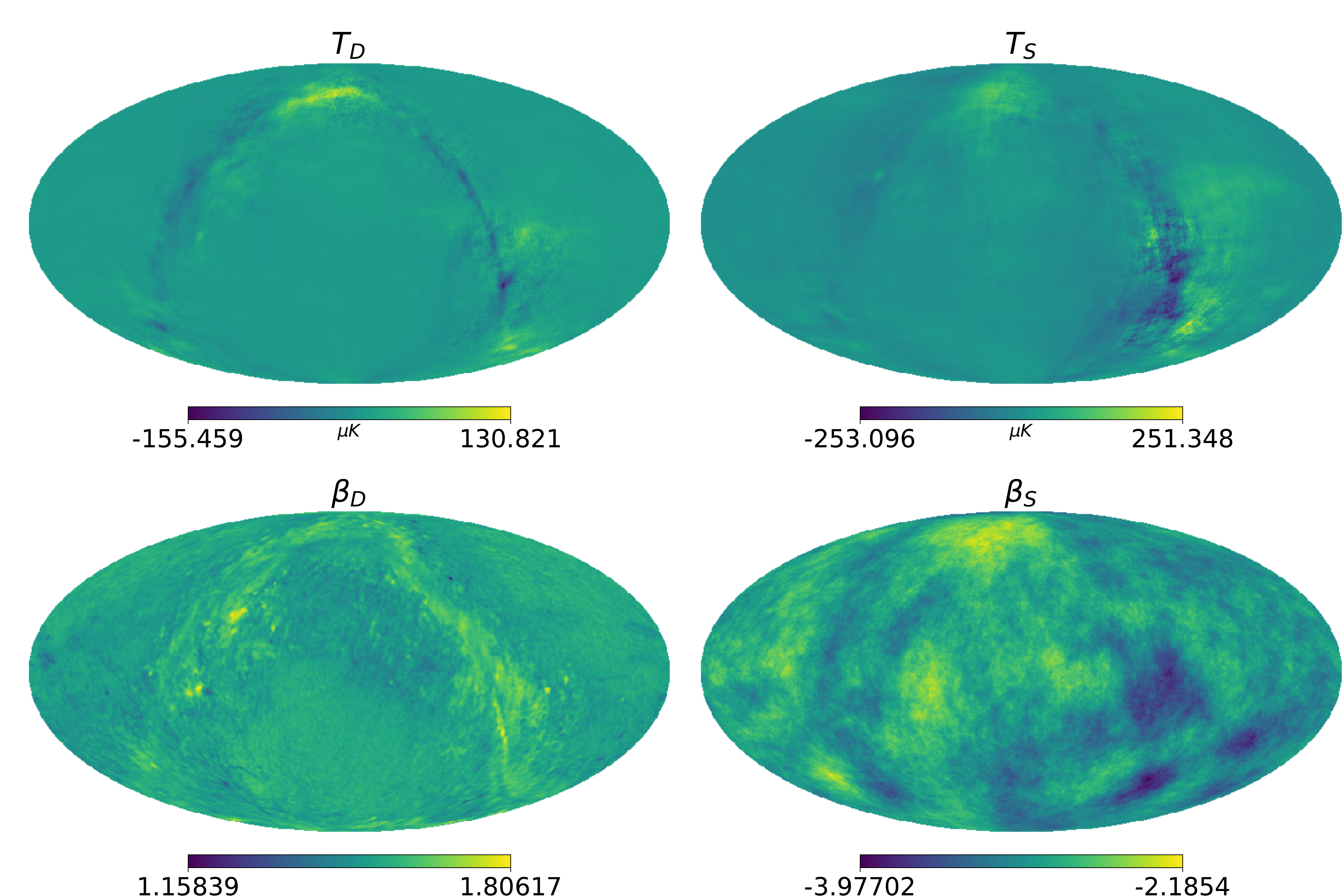}
        \caption{``{\tt d1sm}'' dust and synchrotron polarized amplitude maps (\textit{top row}) and spectral index maps (\textit{bottom row}), in equatorial coordinates.}
        \label{fig:amps_and_betas}
      \end{figure}
      
      Although the level of complexity increases compared to the Gaussian simulations, these simulations contain a comparatively low level of spectral index variation for both dust and synchrotron, compared to the worst-case scenario explored in Gaussian case. For context, the rms fluctuation around the mean of the spectral index maps are
      \begin{equation}\label{eq:sigma pysm} 
        \sigma_{\beta_{\rm D}}=0.04,\hspace{12pt}\sigma_{\beta_{\rm S}}= 0.22.
      \end{equation}
      The foreground amplitude maps used in these simulations reproduce the properties of the power spectrum used to generate the Gaussian realisations described in the previous section (Eq. \ref{eq:gauss_params}) within the footprint use here, and within the scales ($\ell\in[30,300]$) we use in our main analysis. It should be pointed out that, on scales larger than those analyzed here, the foreground power spectra present obvious departures from a perfect power law \cite{2016A&A...594A..10P}, but this does not affect our analysis. Therefore, our model of the scale dependence of foreground amplitudes described in Section \ref{ssec:theory.step2} will not induce any bias on $r$ for these simulations; the bias will arise because of the spatially-varying spectral indices.

    \subsubsection{Additional simulations}\label{sssec:sims.pysm.add}
      To further test the validity of the method, we have generated two additional sets of 20 simulations using other models that introduce different aspects of foreground complexity. In these simulations, the synchrotron SED still corresponds to the \texttt{sm} model described in Section~\ref{ssec:sims.pysm}.

      \begin{itemize}
        \item {\bf Hensley \& Draine (HD).} The thermal dust contribution is modelled using the specifications provided by \cite{2017ApJ...834..134H}, which we label ``HD'' hereafter. This is based on a micro-physical model of dust grains, taking into account the strength of the local radiation field, the grain compositions (carbonaceous, and silicate with varying degrees of iron abundance) and the temperature distribution depending on grain size. In these simulations, the synchrotron SED still corresponds to the \texttt{sm} model described in Section~\ref{ssec:sims.pysm}.
        
        \item {\bf Vansyngel (VS).} An additional suite of 20 simulations was run using the statistical model described in \cite{2017A&A...603A..62V} (labelled ``VS''). This model takes into account the spin orientation of the dust grains, in order to simulate the diffuse polarized emission from galactic dust. The three-dimensional structure of the GMF is described by a finite number of layers (we use $N_{\rm layer}=7$ layers). The coherent component of the magnetic field is the same in all layers, while its turbulent part is generated as a Gaussian random field. Once the direction of the GMF is determined, maps of the $Q$ and $U$ Stokes parameters are generated by scaling the dust intensity map found by \planck{} \cite{2016A&A...594A..10P}. As an additional level of complexity, and in an attempt to describe the three-dimensional distribution of the dust spectral index, we associate each layer with a different Gaussian realization of $\delta\beta_{\rm D}$ with standard deviation $\sigma_{\beta_{\rm D}}=0.13$ (the combined rms variation for 7 layers is $\sigma_{\beta_{\rm D}}\simeq0.35$).
      \end{itemize}

\section{Validation}\label{sec:results}
  We now report on the performance of they hybrid component separation methodology proposed here, as a function of different aspects of foreground complexity. We will present the best-fit and standard deviation on $r$, obtain with the {\sl hybrid} method, and compare it with the results found using the {\sl baseline} $C_\ell$-based method, and with its {\sl moments}-based extension, described in \cite{Azzoni_2021}.
  \begin{figure}
    \centering
    \includegraphics[width=\textwidth]{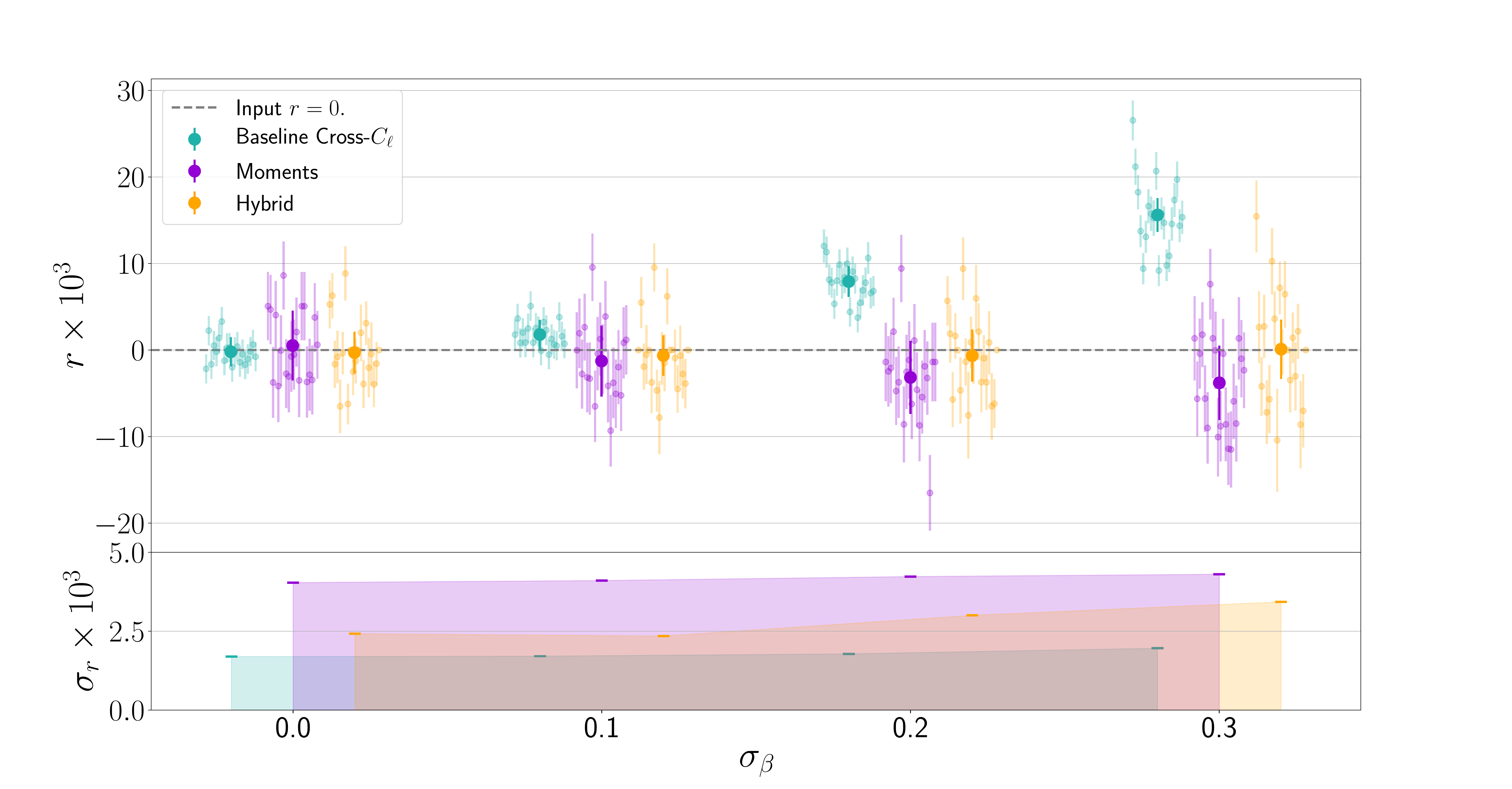}
    \caption{\textit{Upper panel:} Best-fit values of \textit{r} for 20 realizations of the sky calculated at different values of $\sigma_{\beta_c}$ (the same for synchrotron and dust). Results are shown for the baseline cross-$C_\ell$ pipeline assuming constant spectral indices (\textit{turquoise}), adding moments (\textit{purple}) and using the hybrid method (\textit{orange}). The position of each simulation in the $x$ axis is shifted slightly from its true $\sigma_{\beta_c}$ for clarity. The larger, solid dots, at the centre of each $\sigma_{\beta_c}$ value show the mean and standard deviation of each suite of simulations. \textit{Lower panel:} Statistical uncertainty $\sigma_r$ averaged over the twenty realizations in the case of constant spectral indices (\textit{turquoise}), adding moments (\textit{purple}) and using the hybrid method (\textit{orange}). Both the hybrid and the moments method are able to correct the bias on $r$ for all values of $\sigma_{\beta_c}$ considered, at the cost of increased final uncertainties with respect to a model with constant spectral indices (which themselves increase monotonically with $\sigma_{\beta_c}$) .}\label{fig:r_gaussian}
  \end{figure}

  \subsection{Gaussian foreground simulations}\label{ssec:res.gaussian}
  
    \begin{figure}
      \centering
      \includegraphics[width=1.\textwidth]{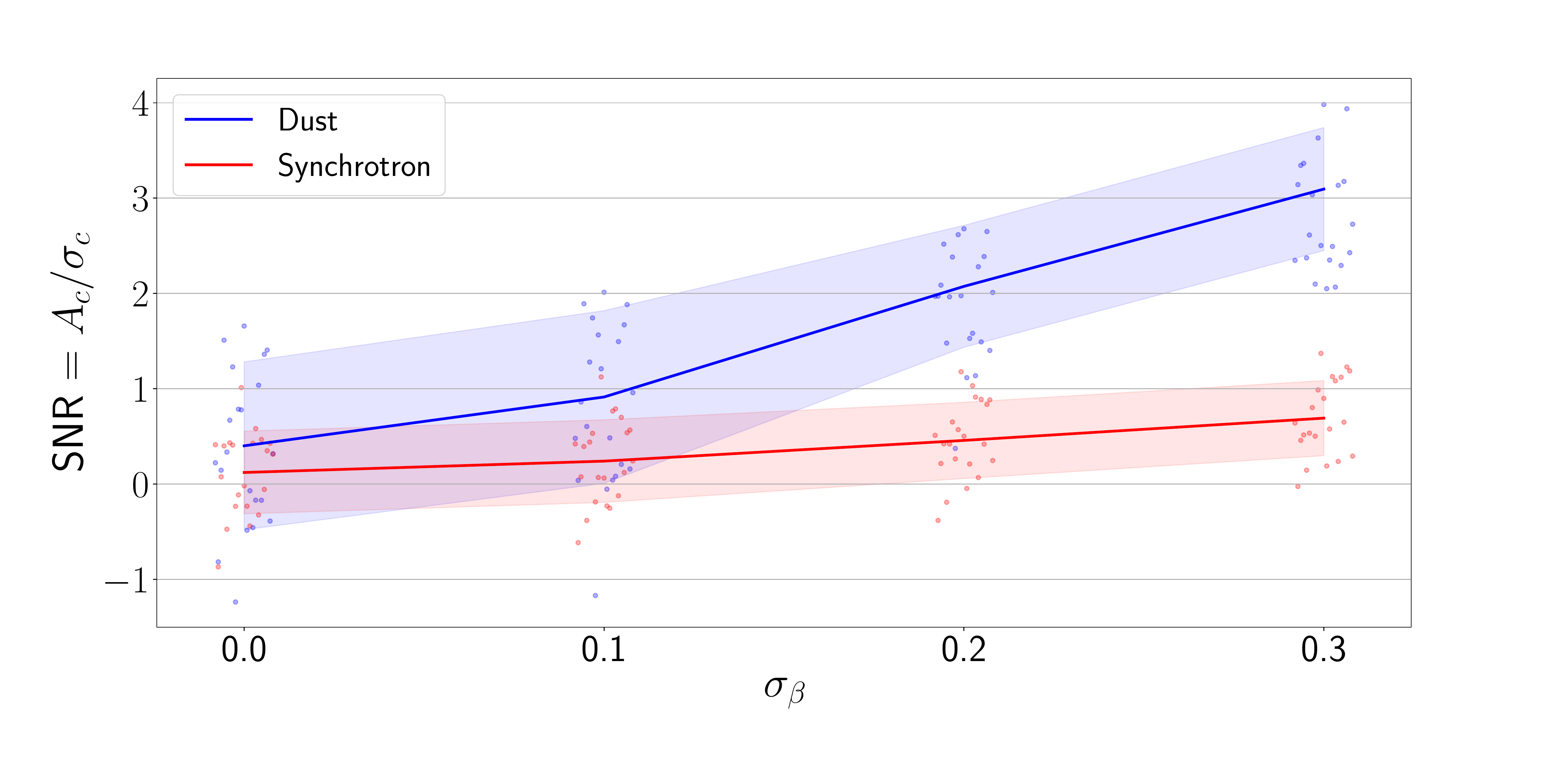}
      \caption{Signal-to-Noise Ratio (SNR) of the foreground amplitudes, dust $A_d^{BB}$ (in blue) and synchrotron $A_s^{BB}$ (in red), using the hybrid method for the Gaussian foreground simulations with increasing spectral index variation ($\sigma_\beta=0-0.3$, the same for dust and synchrotron). For both components, the dots correspond to the SNR of each realization, the line is the mean and the shaded area covers the $1 \sigma$ standard deviation. We find that, at values of $\sigma_\beta > 0.1$, it is possible to significantly detect the effect of the spatial variation of the spectral parameters, particularly the dust spectral index, on the foregrounds multi-frequency power spectra for SO-like sensitivities.}\label{fig:FG_amps}
    \end{figure}

    We use 20 Gaussian simulations with varying values of $\sigma_\beta$, representative of the range allowed by current data \citep{2016A&A...594A..10P}, to explore the impact of increased spatial variability of the spectral indices. 

    The results are shown in Fig. \ref{fig:r_gaussian} as a function of $\sigma_\beta$ (equal for synchrotron and dust). The upper panel shows the best-fit and standard deviation of $\beta$ for each simulation, for the baseline, moments, and hybrid methods in turquoise, purple, and orange, respectively. The darker, larger points, show the overall mean and standard deviation of all simulations for each value of $\sigma_\beta$. The lower panel then shows the standard deviation obtained by each method. The mean measured value of $r$ effectively correspond to the bias we get at different levels of spectral variation, since all the simulations were run with an input $r=0$.

    We find that, as $\sigma_\beta$ increases, the bias on $r$ obtained by the baseline method grows up to $\delta r=0.017$ for the highest value of $\sigma_\beta=0.3$ explored. Since the baseline method achieves uncertainties $\sigma_r\simeq0.002$, this corresponds to a large, $\sim8\sigma$ parameter bias. The hybrid method is then able to correct for this bias in all cases, at the cost of a slight increase in the final parameter uncertainty, with $\sigma_r$ growing to $\sim0.0024$-$0.0033$ (a $\sim20$-$50\%$ increase). The moments expansion method is also able to correct this bias, as was shown in \cite{Azzoni_2021}, although at a slightly higher cost in terms of $\sigma_r$, which rises by $\sim50-70\%$.

    At SO-like sensitivities, we are able to detect the impact of spatially-varying foreground SEDs, particularly in the case of dust. Figure~\ref{fig:FG_amps} shows the Signal-to-Noise Ratio (SNR) of the foreground amplitudes using the hybrid method for the Gaussian foreground simulations with increasing spectral index variation ($\sigma_\beta=0-0.3$, the same for dust and synchrotron). The SNR is calculated for each foreground component ($c$) as the ratio of the amplitude ($A_d^{BB}$ for dust, $A_s^{BB}$ for synchrotron) to the respective $\sigma$ obtained from the posterior distribution. We find that, at values of $\sigma_\beta > 0.1$, it is possible to significantly detect the effect of the spatial variation of the dust spectral index for SO-like sensitivities, while a scatter at the level of $\sigma_\beta\gtrsim0.3$ is required to detect variations in the synchrotron spectral index.
    
  \subsection{Realistic foreground simulations}\label{ssec:res.pysm}
    \begin{figure}
      \centering
      \includegraphics[width=\textwidth]{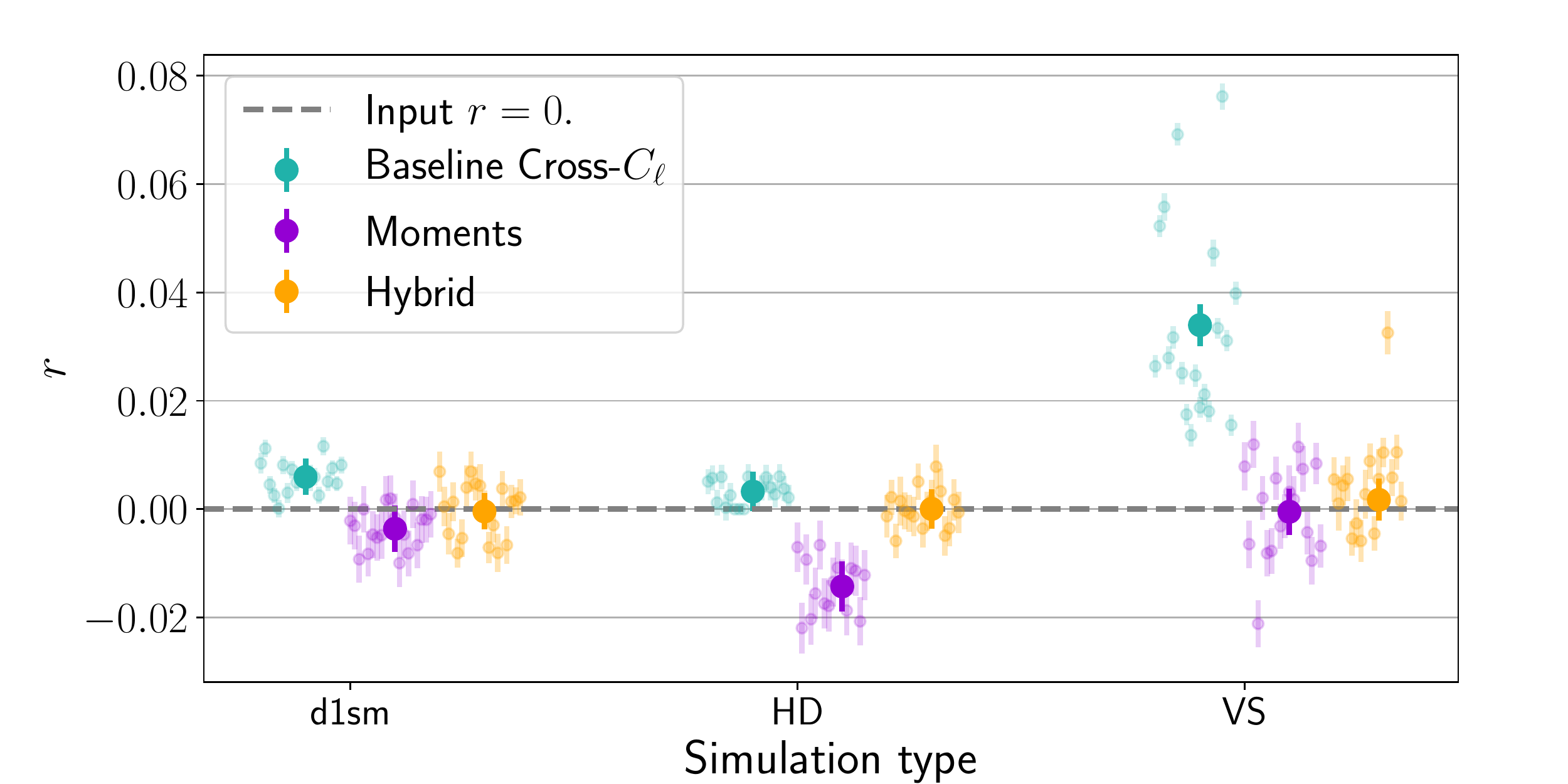}
      \caption{Best-fit values of $r$ for 20 sky simulations for different types of realistic simulations, displayed on the $x$-axis (\texttt{d1sm}, HD, VS). The orange dots show the baseline cross-$C_\ell$ results assuming constant spectral indices, while the purple and orange dots show the results of the hybrid method and the moments method, respectively. The larger, solid dots at the centre of each simulation type show the mean of each suite of simulation. The hybrid method is able to correct the bias on $r$ for all the different types of realistic simulations that appears when assuming constant spectral indices. The moments method recovers unbiased results for the \texttt{d1sm} and VS simulations, but introduces a negative $3 \sigma$ bias in the HD case.}\label{fig:r_d1s1}
    \end{figure}
    Having used the Gaussian simulations to explore the performance of the method as a function of spatial spectral index variability, we now make use of the realistic foreground simulations described in Sections \ref{sssec:sims.pysm.d1sm} and \ref{sssec:sims.pysm.add} to validate the method in the presence of other aspects of foreground complexity. The {\tt d1sm} simulation includes non-Gaussian and statistically inhomogeneous foreground amplitudes, and realistic spectral index maps. The HD simulations make use of a dust spectral law that is not a modified black-body in detail. Finally, the VS simulations account for the three-dimensional variability in the dust spectral index, and include an enhanced level of non-Gaussianity in the dust amplitude. The results are shown in Fig. \ref{fig:r_d1s1} for the three different models (shown in the $x$ axis).

    For {\tt d1sm}, the mean best-fit $r$ from 20 realisations using the baseline method is centered at 0.0059 $\pm$ 0.0023, corresponding to a $\sim2.5\sigma$ bias. With the hybrid method, this reduces to $r=-0.0004 \pm 0.003$. The bias is eliminated at the cost of increasing uncertainties by $\sim 25\%$, as expected from the results found with the Gaussian simulations. Similarly, the moments method recovers unbiased results with a $\sim 40\%$ increase in $\sigma_r$.

    The conclusions are similar when analysing the HD simulations. The different dust SED used in this model introduces a bias at the $1-2\sigma$ level when analysed with the baseline method, obtaining $r=0.0033 \pm 0.002$. The hybrid method corrects this bias (we obtain a mean $r=4.3 \cdot 10^{-5}$), with a widening of the final constraints by $\sim 40\%$. The analysis with the moments method introduces instead a negative $3 \sigma$ bias. This must be due to the foreground bias departing from the theoretical model used here. The two main assumptions are that the variations in $\beta_{\rm D}$ are small enough that the residual spectrum can be described by the Taylor expansion of the modified black-body SED, and that the residuals follow a power-law scale dependence. The additional complexity in the way the non-Gaussian foreground amplitude maps are generated might contribute to either of these causes. It could also be that the simulated data in this case differs significantly from the Gaussian assumption used to construct the covariance matrix and the resulting likelihood is ill behaved.

    The significantly more complex VS simulations introduce a much larger bias on $r$, at the level of $16 \sigma$, when using the baseline method. As in the {\tt d1sm} case, both the hybrid and the moments method eliminate this bias, with an increase in uncertainties by $\sim 30\%$ and $\sim 40\%$ respectively. This is in agreement with the findings of \cite{Azzoni_2021}. Note that one of the key advantages of the hybrid method is that it does not rely on any assumption regarding the Gaussianity of the spectral index variations, unlike the moments method, and therefore it provides more flexibility to handle the additional complexity of the VS model. The impact of this additional complexity, however, does not seem to hamper the moments expansion method significantly by comparison with the hybrid approach explored here.


  \subsection{Larger sky fractions}\label{ssec:res.large_fsky}
    \begin{figure}
      \centering
      \includegraphics[width=0.6\textwidth]{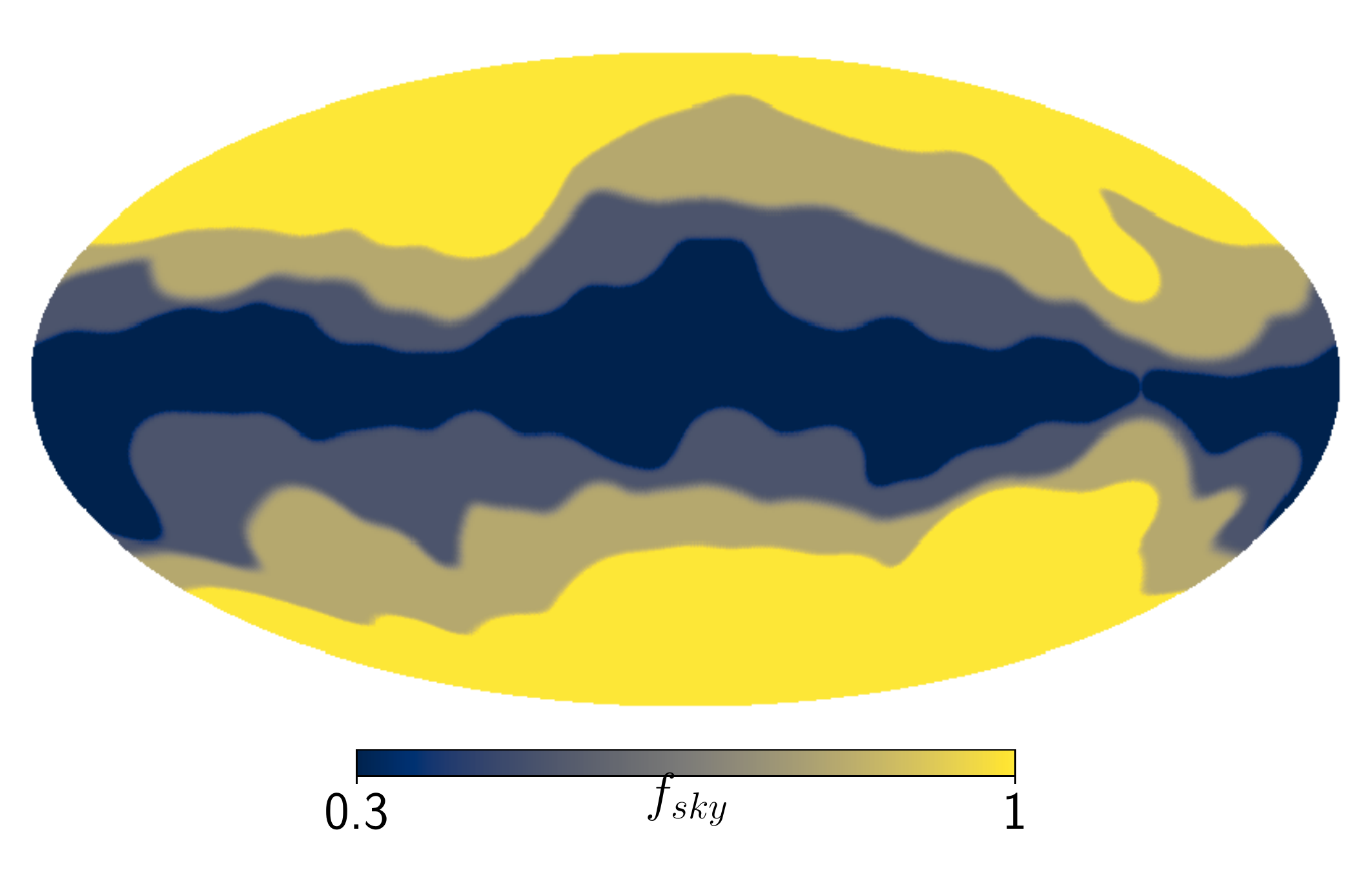}
      \caption{Masks applied to the simulations in Section~\ref{ssec:res.large_fsky}, which exclude the Galactic plane from the estimation of the data. Three masks are created by setting different threshold on the polarization intensity $(P = \sqrt{Q^2 + U^2})$ of the Planck 353 GHz map \cite{2020A&A...643A..42P}, to obtain sky cuts $f_{\rm sky}=$ 0.3 (dark blue), 0.6 (lighter blue) and 0.8 (grey). The yellow area corresponds to the observed footprint in galactic coordinates. }\label{fig:masks_fsky}
    \end{figure}

    The ability of the hybrid method to remove the bulk of the foreground contamination at the map level, without assuming a model for their spatial correlations, or for the correlation between them and the foreground residuals resulting from spatially-varying spectra, could potentially allow us to obtain constraints on sky areas significantly larger than the footprint explored so far. This is relevant in the context of future space missions, such as LiteBIRD \cite{2014JLTP..176..733M, hazumi2019litebird, 2022arXiv220202773L}, or when $B$-mode constraints become cosmic-variance dominated due to imperfect delensing. We have therefore generated and analysed a new set of simulations covering larger sky areas.

\begin{figure}
      \centering
      \includegraphics[width=1.\textwidth]{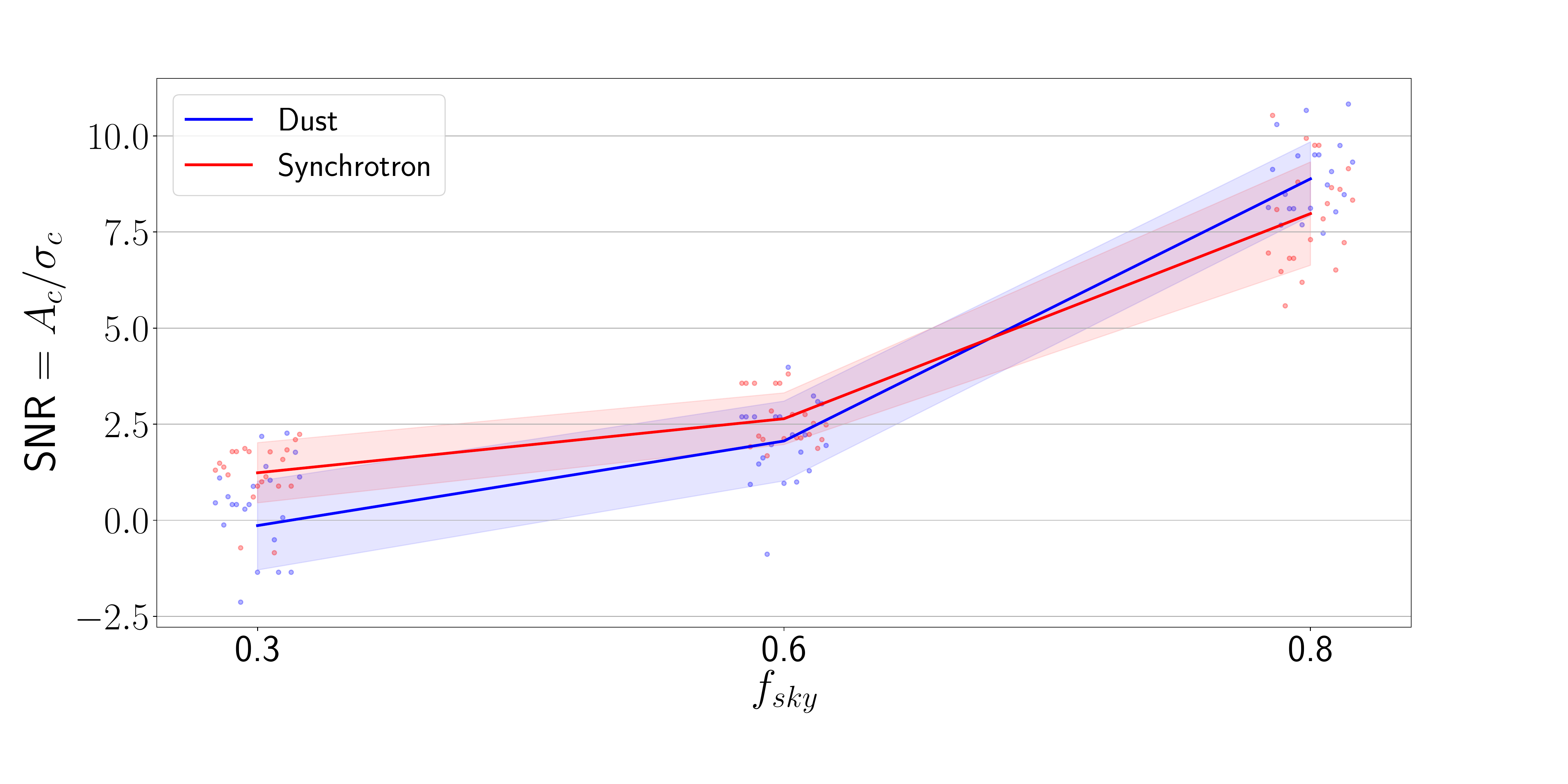}
      \caption{Signal-to-Noise Ratio (SNR) of the foreground amplitudes, dust $A_d^{BB}$ (in blue) and synchrotron $A_s^{BB}$ (in red), using the hybrid method for the \texttt{d1sm} foreground simulations with increasing observed sky fractions ($f_{\rm sky}0.3-0.8$). For both components, the dots correspond to the SNR of each realization, the line is the mean and the shaded area covers the $1 \sigma$ standard deviation. We find that at increasingly larger sky fractions, the foreground contamination is higher, and we can significantly detect the effect of the spatial variation of the spectral parameters on the foregrounds multi-frequency power spectra at $f_{\rm sky} > 30\%$.}\label{fig:SNR_fsky}
    \end{figure}

    In this case, we generate three new sky masks three masks by setting different thresholds on the polarized intensity $(P = \sqrt{Q^2 + U^2})$ of the Planck 353 GHz map \cite{2020A&A...643A..42P}, smoothed on a scale of 1$^\circ$. The thresholds are defined to yield observable sky fractions $f_{\rm sky}=\{0.3,0.6,0.8\}$. The resulting sky masks, avoiding the Galactic plane, are displayed in Fig. \ref{fig:masks_fsky}. 
    These are smoothed with a 10$^\circ$ FWHM beam and and made differentiable with the application of a ``C1'' apodization with a scale of $5^\circ$ to the resulting map \cite{2019MNRAS.484.4127A}. The result is similar to the sky mask used in \cite{planck2018dust}. The larger sky fractions cover regions where foreground contamination is higher, and where spectral index variation is potentially more damaging. For example, in Figure~\ref{fig:SNR_fsky} we find that the mean SNR of the foreground amplitudes recovered with the hybrid method in the {\tt d1sm} model is ${\rm SNR}=0.2, 2.1, 8.9$ (for dust) and ${\rm SNR}=1.2, 2.7, 7.9$ (for synchrotron) in the $f_{\rm sky}=0.3,\,0.6,$ and 0.8 footprints, respectively.
    
    In order to avoid the computational cost of re-calculating the power spectrum covariance matrix from hundreds of simulations for each sky fraction, here we use the approximate formula in Eq.~\ref{eq:knox-covariance}. Note that this approximation is not accurate enough for actual data analysis, which should involve extensive simulations, but it serves the needs of this forecasting exercise, as illustrated in \cite{2019JCAP...02..056A}. The power spectra introduced in the formula contain both signal and noise contributions. Although the approximation accounts for the impact of incomplete sky observations in reducing the number of available independent modes, it does not account for the effects of mode-mode coupling, or the mixing of $E$ and $B$ modes. To get around this limitation, we have also modified the simulations used for this part of the analysis. We remove any $E$-to-$B$ leakage from the simulations by hand, by transforming the original $Q$ and $U$ maps into $E$ and $B$, setting the $E$-mode component to zero, and transforming back to $Q$ and $U$. Although this degrades the level of realism used in the previous sections, it allows us to account for the main impact of foreground complexity: the large dynamic range of foregrounds on large scale areas, and the departure from perfect power-law behaviour on large scales. For simplicity, in these simulations we assume the same noise power spectrum used before (Eq. \ref{eq:nl}), with a flat sky coverage (i.e. the noise properties are homogeneous throughout the observed footprint).
    
    \begin{figure}
      \centering
      \includegraphics[width=\textwidth]{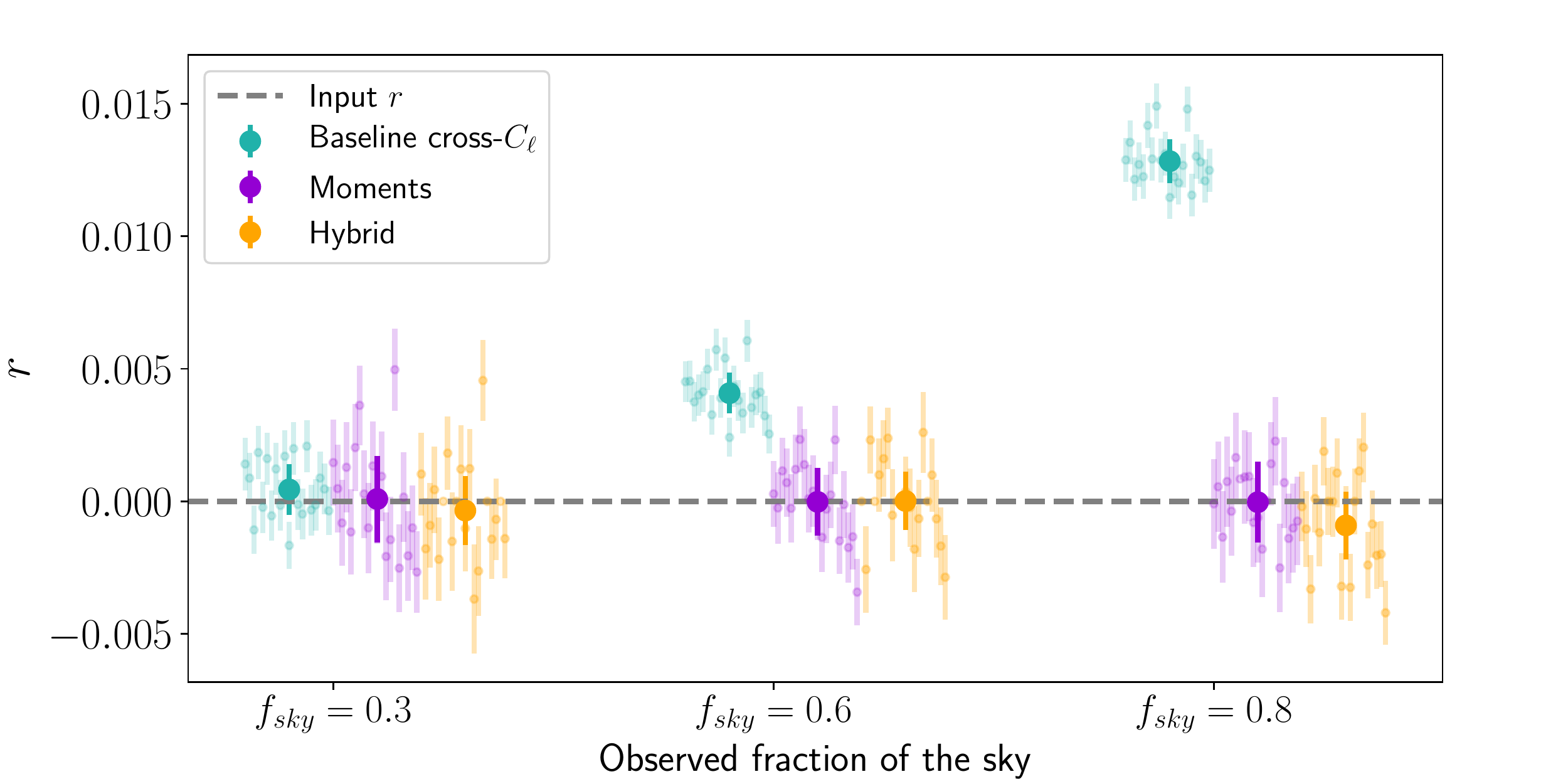}
      \caption{Constraints on $r$ obtained by the ``baseline'', ``moments'', and ``hybrid'' methods (in turquoise, purple and orange), as a function of sky fraction, for the {\tt d1sm} foreground model. With sky areas wider than 30\% ($f_{\rm sky} > 0.3$), the baseline pipeline yields biased results. Both the hybrid method and the moments method are able to eliminate this bias, with an increase in $\sigma_r$ by $30-40\%$ (hybrid) and by $50-80\%$ (moments) with respect to the baseline constraints.}\label{fig:r_fsky}
    \end{figure}

    \begin{figure}
      \centering
      \includegraphics[width=\textwidth]{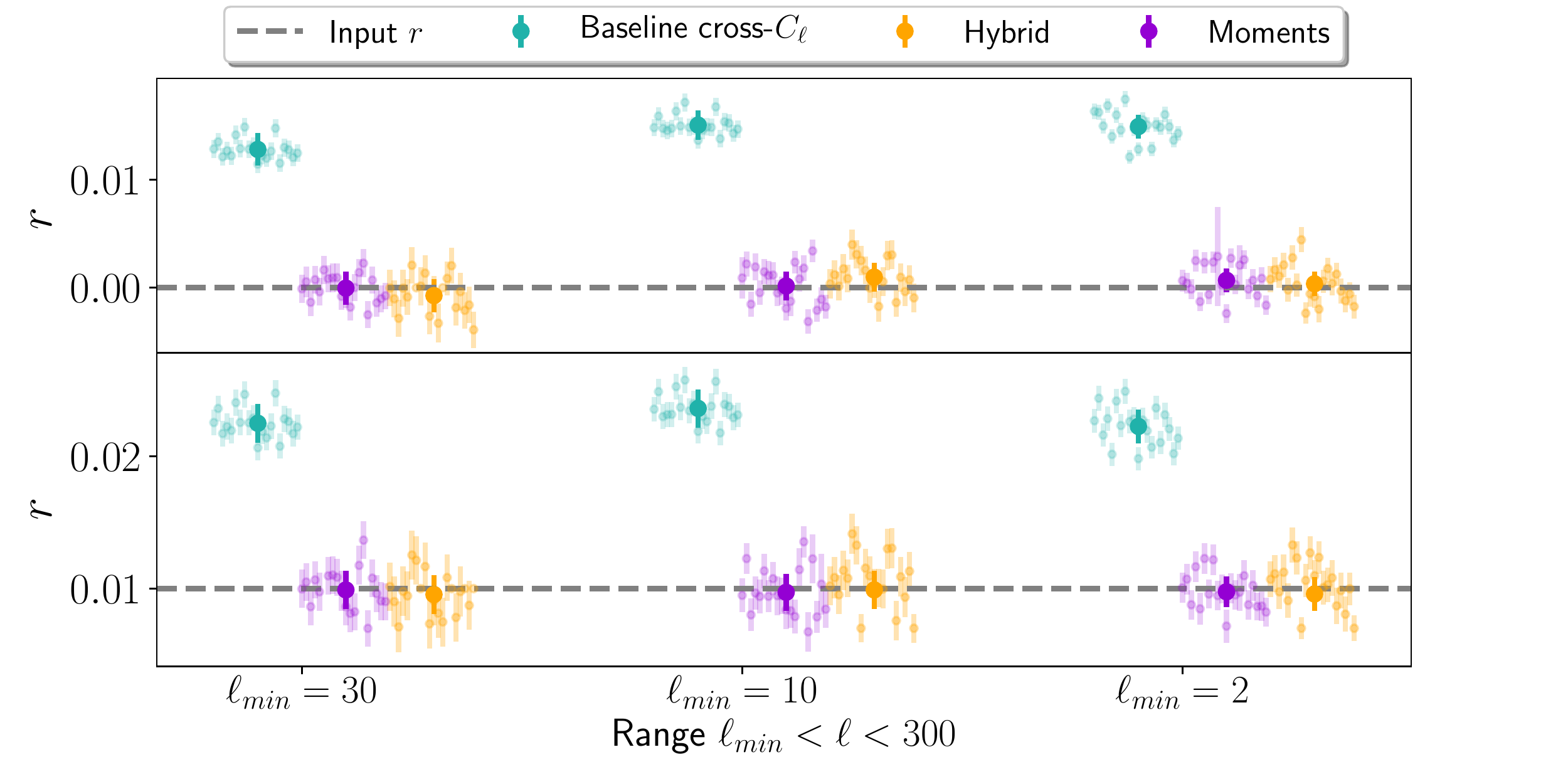}
      \caption{Best-fit values of $r$ for 20 sky simulations observing 80\% of the sky $(f_{\rm sky} = 0.8)$ across different ranges of angular scales (from $\ell_{\rm min}$ up to $\ell = 300$) for input $r=0$ (\textit{upper panel}) and $r=0.01$ (\textit{lower panel}). The $x-$axis displays increasing values of angular scales (lower $\ell_{\rm min}$). The red dots show the baseline cross-$C_\ell$ results, while the green dots show the results of the hybrid method. The larger, solid dots at the centre of each suite of simulation show the mean across simulations. The hybrid method is able to correct the bias on $r$ even at higher angular scales down to $\ell_{\rm min}=2$.}\label{fig:r_lmin}
    \end{figure}

    Fig. \ref{fig:r_fsky} shows the constraints on $r$ found with the ``baseline'', ``moments'', and ``hybrid'' methods, as a function of sky fraction, for the {\tt d1sm} foreground model. Over the 30\% footprint, all methods obtain unbiased constraints on $r$. This, however, changes rapidly as we increase the sky area. At $f_{\rm sky}=0.6$ and 0.8, the baseline method yields $r=0.004 \pm 0.0008$ and $r=0.013 \pm 0.0009$, corresponding to a $\sim 5\sigma$ and a $\sim 16\sigma$ bias, respectively. Remarkably, the hybrid method is able to absorb this bias in all cases ($r= 0.00001 \pm 0.0011$ and $-0.0007 \pm  0.0012$ for 60\% and 80\% $f_{\rm sky}$ respectively), even when using 80\% of the sky. The validity of this result, obviously depends on the level of realism of the foreground simulations used here, which may not be sufficiently accurate close to the Galactic plane. However, the result highlights the robustness of the method to high levels of foreground contamination, in comparison with standard approaches. Similarly, the moments method is able to recover unbiased results, recovering $r= 0.0 \pm 0.0013$ and $-0.0001 \pm  0.0015$ for 60\% and 80\% $f_{\rm sky}$, but with a $20-30\%$ increase in $\sigma_r$ compared to the hybrid method constraints.

    The availability of a larger sky area allows us to constrain primordial $B$-modes using larger scales. Fig. \ref{fig:r_lmin} shows the constraints obtained when using the $f_{\rm sky}=0.8$ footprint as a function of the minimum multipole, $\ell_{\rm min}$, included in the analysis. The two panels in this figure show the results for simulations with an input $r=0$ (top), and for $r=0.01$ (bottom), which thus allows us to study the performance of the method in the presence of a detectable $B$-mode signal. 
    For our default $\ell_{\rm min}=30$, we obtain unbiased constraints on $r$ with the hybrid approach in both cases ($r=-0.0007\pm 0.0012$ and $r=0.009\pm 0.0013$), whereas the baseline approach recovers a biased $r$ at the same level as that displayed in Fig. \ref{fig:r_fsky}. 
    When including larger angular scales ($\ell_{\rm min} < 30$), we use the HL non-Gaussian likelihood \cite{2008PhRvD..77j3013H}, as described in Sec.~\ref{sssec:theory.step2.cllike}.  When extending the scale range to $\ell_{\rm min}=10$, the hybrid method obtains $r=0.0005 \pm 0.0013$ (input $r=0$) and $r=0.0099\pm 0.0013$ (input $r=0.01$).
    When including even larger scales, down to $\ell_{\rm min}=2$, the hybrid method recovers again unbiased results, i.e. $r=0.0002 \pm 0.0011$ (input $r=0$) and $r=0.0096\pm 0.0012$ (input $r=0.01$), compared to the $19\sigma$ bias introduced by the baseline cross-$C_\ell$ pipeline. 
    Similarly, the moments method recovers the input $r$ value on all scales, with a $\sim 10\%$ increase in $\sigma_r$ compared to the hybrid method, and introducing a small negative $< 0.2 \sigma$ bias.

\section{Conclusion}\label{sec:conclusion}
  Detecting the signature of inflation in the CMB polarized $B$-mode signal is one of the most compelling goals of cosmology, where a strict upper constraint on the large scales amplitudes has the potential to further support or disprove various inflationary models \cite{2014JCAP...03..039M}. 

  Many sources of systematic uncertainty contaminate the faint primordial $B-$modes with the creation of ``spurious'' $B-$mode signal. Among these, the most significant astrophysical source is coming from Galactic polarized foregrounds \cite{2014PhRvL.112x1101B}. 
  Therefore, one of the most important steps in the analysis of CMB $B$-mode data is the separation of the multi-frequency data into its different components. In principle, the separation of CMB radiation and foregrounds can be performed either in real space, dealing with maps of the $Q/U$ Stokes parameters, or in Fourier space, processing the $E/B$ modes power spectra. Different CMB experiments have followed either approach, revealing their advantages and caveats. Whilst achieving satisfactory results on simulated data, current methods are facing additional challenges put forth by the increasing sensitivity of the next-generation CMB experiments, where the precision of the different approaches will become crucial. 

  Here, we investigate a hybrid approach that combines characteristics from both map-based and $C_\ell$-based studies. In the first step of the method, we determine foreground amplitudes at the map level without making any assumptions about their statistical correlations and assuming spatially uniform spectral characteristics. We next take this spatially-homogeneous foreground contribution away from the original data, and the residual foreground contamination (due to inaccurate subtraction or spatial SED fluctuations) is modelled at the power spectrum level. Therefore, only at this stage we This allows us to retain the same $C_\ell$-level assumptions, imposing a model for the scale dependence of the foreground amplitudes, while also limiting the degradation in the final constraints brought on by spectral index variations, by applying those assumptions to the smaller foreground residuals (instead of the full foreground spectra) and retaining information about their frequency dependence. Therefore this method allows us to overcome the limitations imposed by $C_\ell$ approaches (Gaussianity, power spectrum form), limiting the number of parameters needed to model the smaller foreground residuals. 

  We have applied this method to simulated data including realistic foreground models and instrumental noise. We have shown that we are able to obtain unbiased constraints on $r$ in the presence of realistic (and often pessimistic) foreground complexity. We compare this with the results of a ``baseline'' fiducial cross-$C_\ell$ analysis (analogue to the method used to derive the current state-of-the-art constraints on the tensor-to-scalar-ratio $r$ from $B$-modes \cite{2016PhRvL.116c1302B}), and with the extended ``moments'' approach described in \cite{Azzoni_2021}, which is able to further mitigate contamination from foreground residuals. Generally, we find that the hybrid method outperforms the baseline one in terms of bias, and the moments method in terms of $\sigma_r$ (and bias for some of the most complicated foreground models). 

  In the simpler foreground scenario, which includes no spatial variation of the spectral parameters (corresponding to the Gaussian simulations with $\sigma_\beta = 0$), we obtain unbiased tensor-to-scalar ratio $r$ with all the three pipelines, with a respective increase in $\sigma_r$ by $\sim 30\%$ (moments) and $<10\%$ (hybrid) compared to the baseline approach. In Section~\ref{ssec:res.gaussian}, we find that increasing spectral index variations ($\sigma_\beta=0.1-0.3$) can introduce a bias on $r$ of up to $8\sigma$ with the baseline method, which both the hybrid and the moments methods are able to correct for at the cost of an increase in statistical uncertainty (the former by $\sim 20-50 \%$, the latter by $\sim 50-70\%$).

  We have also explored more realistic foreground simulations based on various models proposed in the literature (Section \ref{ssec:res.pysm}) with different levels of complexity. Overall, we find that the hybrid method is able to correct the bias on $r$ for all the different types of realistic simulations that appears when assuming constant spectral indices, at the cost of a limited $\sigma_r$ increase by $< 30 \%$. The moments analysis recovers unbiased results for most of these simulations. However, the reliance of the moments method on simplifying assumptions (Gaussianity of the spectral index variations, uncorrelation among spectral indices and foreground amplitudes) limits its ability to recover unbiased results for one type of complex foreground model explored here. 

  Unlike the moments method, the hybrid approach has the advantage of not relying on any assumptions about the shape of the spectral index fluctuations, and proceeds by removing the bulk of the foreground contamination at the map level, without assuming a model for their spatial correlations, or for the correlation between them and the foreground residuals resulting from spatially-varying spectra. This allows us to obtain constraints on increasingly larger sky areas and at larger angular scales in Section \ref{ssec:res.large_fsky}. Here we find that the results remain largely unbiased, even for experiments covering $\sim80\%$ of the sky, suggesting the potential application of this method to future satellite missions such as LiteBIRD.

\acknowledgments
  We thank Carlo Baccigalupi, Kevin Wolz, Davide Poletti for many helpful discussions. SA is funded by a Kavli/IPMU PhD Studentship. MHA acknowledges support from the Beecroft Trust and Dennis Sciama Junior Research Fellowship at Wolfson College. DA acknowledges support from the Beecroft Trust, and from the Science and Technology Facilities Council through an Ernest Rutherford Fellowship, grant reference ST/P004474. This publication arises from research funded by the John Fell Oxford University Press Research Fund. Some of the results in this paper have been derived using the {\tt HEALPix} package \cite{2005ApJ...622..759G}. The contour plots were generated using the {\tt GetDist} package \cite{2019arXiv191013970L}. We made extensive use of the {\tt numpy} \citep{OliphantNumPy, 2011CSE....13b..22V}, {\tt scipy} \citep{2020NatMe..17..261V}, {\tt healpy} \citep{2019JOSS....4.1298Z}, and {\tt matplotlib} \citep{2007CSE.....9...90H} python packages.

\appendix

\bibliography{bibliography}

\end{document}